\documentclass[journal]{IEEEtran}

\pdfoutput=1
\usepackage{outline}
\usepackage{pmgraph}
\usepackage[normalem]{ulem}
\usepackage[utf8]{inputenc}
\usepackage{amssymb}
\usepackage{hyperref}
\usepackage{amsmath}
\usepackage{graphicx}
\usepackage{times}
\usepackage{xcolor}
\usepackage{xspace}
\usepackage[colorinlistoftodos]{todonotes} 
\usepackage{cite}
\usepackage{bm} 
\usepackage{url}

\graphicspath{{fig/}} 

\usepackage{tikz}
\usetikzlibrary{spy}
\usepackage{algpseudocode}
\usepackage{algorithm}
\usepackage{mathrsfs}

\newcommand{\add}[1] {\textcolor{black}{#1}} 
\newcommand{\addnew}[1] {\textcolor{black}{#1}} 

\def\QED{~\rule[-1pt]{5pt}{5pt}\par\medskip}

\long\def\comment#1{} 
\newcommand{\term}[1] {\emph{#1}}
\newcommand{\reg} {\xmath{\beta}}

\DeclareMathOperator*{\argminop}{arg\,min} 
\def\argmin#1{\argminop_{#1}}

\newcommand{\rank}{\operatorname{rank}}

\newcommand{\xmath}[1] {\ensuremath{#1}\xspace}
\newcommand{\blmath}[1] {\xmath{\bm{#1}}}
\newcommand{\A}{\blmath{A}}
\newcommand{\B}{\blmath{B}}

\newcommand{\D}{\blmath{D}}
\newcommand{\E}{\blmath{E}}

\newcommand{\R}{\blmath{R}}
\newcommand{\T}{\blmath{T}}

\newcommand{\W}{\blmath{W}}

\newcommand{\Z}{\blmath{Z}}
\renewcommand{\P}{\blmath{P}} 
\renewcommand{\H}{\blmath{H}}
\newcommand{\G}{\blmath{G}}
\newcommand{\x}{\blmath{x}}
\newcommand{\xh}{\xmath{\hat{\x}}}
\newcommand{\y}{\blmath{y}}
\newcommand{\z}{\blmath{z}}
\newcommand{\zh}{\xmath{\hat{\z}}}

\newcommand{\ie}{\text{i.e.}}
\newcommand{\eg}{\text{e.g.}}

\newcommand{\norm}[1] {\xmath{\left\| #1 \right\|}}
\newcommand{\normi}[1] {\xmath{\left\| #1 \right\|_1}}
\newcommand{\normii}[1] {\xmath{\left\| #1 \right\|_2}}

\newcommand{\Db}{{\blmath D}}
\newcommand{\Eb}{{\blmath E}}

\newcommand{\mb}{{\blmath m}}

\newcommand{\Rc}{\mathcal{R}}

\newcommand{\Phib}{{\boldsymbol {\Phi}}}
\newcommand{\Psib}{{\boldsymbol {\Psi}}}

\newcommand{\Sigmab}{{\boldsymbol {\Sigma}}}
\newcommand{\Rd}{{\mathbb R}}
\newcommand{\Cd}{{\mathbb C}}

\newcommand{\phib}{{\boldsymbol{\phi}}}
\newcommand{\psib}{{\boldsymbol{\psi}}}

\newcommand{\Ybc}{{\boldsymbol{\mathcal Y}}}

\newcommand{\Zbc}{{\boldsymbol{\mathcal Z}}}

\newcommand{\thetab}{{\boldsymbol {\theta}}}

\newcommand{\Xbc}{{\boldsymbol{\mathcal X}}}

\newcommand{\Ec}{{{\mathcal E}}}
\newcommand{\Dc}{{{\mathcal D}}}

\newcommand{\hank}{\boldsymbol{\mathbb{H}}}

\newcommand{\beq}{\begin{equation}}
\newcommand{\eeq}{\end{equation}}
\newcommand{\beqa}{\begin{eqnarray}}
\newcommand{\eeqa}{\end{eqnarray}}
\usepackage{subfigure} 

\graphicspath{{fig/}}

\begin{document}
\title{Image Reconstruction: From Sparsity to Data-adaptive Methods and Machine Learning}


\author{Saiprasad~Ravishankar,~\IEEEmembership{Member,~IEEE,}%
~Jong~Chul~Ye,~\IEEEmembership{Senior Member,~IEEE,}%
~and~Jeffrey~A.~Fessler,~\IEEEmembership{Fellow,~IEEE}
\thanks{DOI: 10.1109/JPROC.2019.2936204. Copyright (c) 2019 IEEE. Personal use of this material is permitted. Permission from IEEE must be obtained for all other uses, in any current or future media, including reprinting/republishing this material for advertising or promotional purposes, creating new collective works, for resale or redistribution to servers or lists, or reuse of any copyrighted component of this work in other works.
}
\thanks{
J.~C.~Ye was supported in part by the
National Research Foundation of Korea grant NRF-2016R1A2B3008104; and
J.~A.~Fessler was supported in part by National Institutes of Health (NIH) grants
R01 CA214981,
R01 EB023618,
U01 EB018753, and
U01 EB026977.
}
\thanks{
S.~Ravishankar is with the Departments of Computational Mathematics, Science and Engineering,
and Biomedical Engineering at Michigan State University, East Lansing, MI, 48824 USA (email: ravisha3@msu.edu).
J.~C.~Ye is with the Department of Bio and Brain Engineering and Department of Mathematical Sciences at the Korea Advanced Institute of Science \& Technology (KAIST), Daejeon, South Korea (email: jong.ye@kaist.ac.kr). 
J.~A.~Fessler is with the Department of Electrical Engineering and Computer Science,
University of Michigan, Ann Arbor, MI, 48109 USA (email: fessler@umich.edu).}
}
\maketitle 


\begin{abstract}
The field of \add{medical} image reconstruction
has
\add{seen roughly four types}
of methods.
The first \add{type tended to be} 
analytical methods,
such as filtered back-projection (FBP) for X-ray computed tomography (CT)
and the inverse Fourier transform for magnetic resonance imaging (MRI),
based on simple mathematical models for the imaging systems.
These methods are typically fast,
but have suboptimal properties
such as poor resolution-noise trade-off for CT.
A second \add{type is}
iterative reconstruction methods
based on more complete models for the imaging system physics and,
where appropriate, models for the sensor statistics.
These iterative methods improved image quality by reducing noise and artifacts.
The FDA-approved methods among these have been based on relatively simple regularization models.
\add{A third type}
of methods has been designed to accommodate modified data acquisition methods,
such as reduced sampling in MRI and CT
to reduce scan time or radiation dose.
These methods typically involve mathematical image models
involving assumptions such as \term{sparsity} or \emph{low-rank}.
\add{A fourth type}
of methods
replaces mathematically designed models of signals \add{and systems}
with \emph{data-driven} or \emph{adaptive} models inspired by the field of \emph{machine learning}.
This paper
\addnew{focuses on the two most recent trends
in \add{medical} image reconstruction:} 
methods based on sparsity or low-rank models,
and data-driven methods based on machine learning techniques.
\end{abstract}


\section{Introduction}
\label{sec:introduction}

Various \add{medical} imaging modalities are popular in clinical practice,
such as magnetic resonance imaging (MRI), X-ray computed tomography (CT),
positron-emission tomography (PET), single-photon emission computed tomography (SPECT), etc.
These modalities help image various biological and anatomical structures
and physiological functions, and aid in medical diagnosis and treatment.
Ensuring high quality 
images reconstructed from limited or corrupted (e.g., noisy) measurements
such as subsampled data in MRI (reducing acquisition time)
or low-dose or sparse-view data in CT
(reducing patient radiation exposure)
has been a popular area of research
and holds high value 
in improving clinical throughput and patient experience.
This paper 
reviews
some of the major recent advances in the field of image reconstruction,
focusing on methods that use sparsity, low-rankness, and machine learning.
\addnew{We focus partly on PET, SPECT, CT, and MRI examples,
but the general methods
can be useful for other modalities,
both medical and non-medical.
This paper is part of a special issue
that focuses on sparsity and machine learning in medical imaging.
Other papers in this issue emphasize other modalities.
}

\subsection{\add{Types of Image Reconstruction Methods}}

Image reconstruction methods
have undergone significant advances
over the past few decades,
\add{
with different paths for
various modalities.
}
These advances can be broadly grouped in four
\add{categories of methods}.
\add{The} first 
\add{category consists of}
analytical and algebraic methods.
These methods include the classical filtered back-projection (FBP) methods for X-ray CT
(e.g., Feldkamp-Davis-Kress or FDK method \cite{feldkamp:84:pcb})
and the inverse Fast Fourier transform
and extensions such as the Nonuniform Fast Fourier Transform (NUFFT)%
~\cite{fessler:03:nff,defrancesco:04:enb} 
for MRI and CT.
These methods are based on relatively simple mathematical models of the imaging systems,
and although they have efficient \add{and} fast implementations,
they suffer from suboptimal properties such as poor resolution-noise trade-off for CT.

A second \add{category} of reconstruction methods
involves iterative reconstruction algorithms
that are based on more sophisticated models
for the imaging system's physics and models for sensor and noise statistics.
Often called model-based image reconstruction (MBIR) methods
or statistical image reconstruction (SIR) methods,
these schemes iteratively estimate the unknown image
based on the system (physical or forward) model, measurement statistical model,
and assumed prior information about the underlying object
\cite{
sauer:93:alu,
thibault:06:arf}.
For example, minimizing penalized weighted-least squares (PWLS) cost functions
has been popular in many modalities including PET and X-ray CT,
and these costs include a statistically weighted quadratic data-fidelity term
(capturing the imaging forward model and noise \term{variance})
and a penalty term called a regularizer
that models the prior information about the object
\add{\cite{depierro:93:otr}}.
These iterative reconstruction methods improve image quality
by reducing noise and artifacts. 
In MRI, parallel data acquisition methods (P-MRI)
exploit the diversity of multiple receiver coils
to acquire fewer Fourier or k-space samples \cite{pMRI-Survey}.
Today, P-MRI acquisition is used widely 
in commercial systems,
and MBIR-type methods in this case
include those based on coil sensitivity encoding (SENSE) \cite{pMRI-Survey}, etc.
The iterative medical image reconstruction methods approved
by the U.S. food and drug administration (FDA)
for SPECT, PET, and X-ray CT
have been based on relatively simple regularization models.

A third \add{category} of reconstruction methods
accommodate modified data acquisition methods
such as reduced sampling in MRI and CT
to significantly reduce scan time and/or radiation dose.
Compressed sensing (CS) techniques
\cite{feng96a,BreFen-C96c,donoho2006compressed,emmanuel2004robust,ye2002self}
have been particularly popular among this class of methods
(leading to journal special issues
\add{\cite{baraniuk:10:aos}}
\cite{wangbresntzi11}).
These methods have been so beneficial for MRI~\cite{lustig2007sparse,CSMRIreview}
that they recently got FDA approval~\cite{fda:17:ge,fda:17:siemens,fda:18:philips}.
CS theory predicts the recovery of images from far fewer measurements
than the number of unknowns,
provided that the image is sparse in a transform domain or dictionary,
and the acquisition or sampling procedure is appropriately incoherent with the transform.
Since MR acquisition in Fourier or k-space occurs sequentially over time,
making it a relatively slow modality,
CS for MRI can enable quicker acquisition by collecting fewer k-space samples.
However, the reduced sampling time comes at the cost
of slower, nonlinear, iterative reconstruction.
The methods for reconstruction from limited data
typically exploit mathematical image models
based on \emph{sparsity} or \emph{low-rank}, etc.
In particular, CS-based MRI methods often use variable density random sampling techniques
to acquire the data
and use sparsifying transforms such as wavelets, finite difference operators
(via total variation (TV) penalty), contourlets, etc.,
for reconstruction~\cite{lustig2007sparse,qu2010iterative}.
Research \add{about such methods}
also focused on developing new theory and guarantees
for sampling and reconstruction from limited data~\cite{adchanpoonrom13},
and on new optimization algorithms
for reconstruction with good convergence rates~%
\cite{
kim:15:cos}.

A fourth \add{category} of image reconstruction methods
replaces mathematically designed models of images and processes
with \emph{data-driven} or \emph{adaptive} models
inspired by the field of \emph{machine learning}.
Such models
(e.g., synthesis dictionaries~\cite{Aharon2006},
sparsifying transforms~\cite{sai2013tl}, tensor models, etc.)
can be learned in various ways
such as by 
using training
datasets~\cite{xu:12:ldx,zheng:18:pua},
or even learned jointly with the reconstruction%
~\cite{sai2011dlmri,xu:12:ldx,lingaljacob13,sravTCI1},
a setting called 
model-blind reconstruction or blind compressed sensing (BCS)~\cite{eldar2011bcs}. 
While most of these methods perform \emph{offline} reconstruction
(where the reconstruction is performed once all the measurements are collected),
recent works show that the models can also be learned
in a time-sequential or \emph{online} manner
from streaming measurements to reconstruct dynamic objects%
~\cite{mardanimateosgiannakis15,briansairajjeff18}.
The learning can be done in an unsupervised manner
employing model-based and surrogate cost functions,
or the reconstruction algorithms
(such as deep convolutional neural networks (CNNs))
can be trained in a supervised manner
to minimize the error in reconstructing training datasets
that typically consist of pairs of ground truth and \addnew{undersampled data%
~\cite{schlemper18,leeye18,ravchfess17,wang:16:apo,shan:19:cpo}.}
These 
learning-based reconstruction methods
form a very active field of research
with numerous conference special sessions and special journal issues
devoted to the topic~\cite{wang:18:iri}.

\add{The categories above
are not a strict chronology;
for example,
NN methods were investigated for image reconstruction
as early as 1991
\cite{floyd:91:aan}, 
\addnew{and for MR spectroscopy soon thereafter
\cite{venkataraman:94:ann}},
and some of the earliest methods for X-ray CT were iterative.
}

\subsection{Focus and Outline of This Paper}

This paper reviews the progress in medical image reconstruction,
focusing on the two most recent trends:
methods based on sparsity using analytical models,
and low-rank models and extensions that combine sparsity and low-rank, etc.;
and data-driven models and approaches exploiting machine learning.
Some of the mathematical underpinnings
and connections between different models and their pros and cons are also discussed.

The paper is organized as follows.
Section~\ref{sec:approaches} describes early image reconstruction approaches,
especially those used in current clinical systems.
Sections~\ref{sec:sparsity} and \ref{sec:lowrank} describe
sparsity and low-rank based approaches
for image reconstruction.
Section~\ref{sec:datadriven} surveys 
the advances in data-driven image models and related machine learning approaches for image reconstruction.
Among the learning-based methods,
techniques that learn image models using model-based cost functions
from training data,
or on-the-fly from measurements are discussed,
followed by recent methods relying on supervised learning of models for reconstruction,
typically from datasets of high quality images and their corrupted versions.
Section~\ref{sec:deepmodel} reviews the very recent works using learned convolutional neural networks (a.k.a. deep learning) for image reconstruction.
Section~\ref{sec:open} discusses some of the current challenges and open questions
in image reconstruction and outlines future directions for the field.
Section~\ref{sec:conclusion} concludes this review paper.

\section{\addnew{Iterative Reconstruction \addnew{Used Clinically}}}
\label{sec:approaches}

This section focuses on
\addnew{some of}
the iterative MBIR methods
that are in routine clinical use currently,
and relates the models used in those systems
to the sparsity models
used in the contemporary literature.
As mentioned in the introduction,
MBIR methods have been used routinely 
for many years
in commercial SPECT, PET and CT systems.
Early publications on MBIR methods
tended to focus on
\addnew{mathematical Bayesian models}.
In contrast,
recent data-driven methods
are based on empirical distributions
from training data,
as discussed later in the paper.
The dominant Bayesian approach
for reconstructing an image \xh
from data \y
was the maximum \emph{a posteriori} (MAP)
approach of finding the maximizer
of the posterior
$p(\x|\y)$.
By Bayes rule,
the MAP approach is equivalent to
\begin{equation}
\xh = \argmin{\x} 
\left( -\log p(\y|\x) - \log p(\x) \right)
\label{e,xh,map}
,\end{equation}
where
$-\log p(\y|\x)$
denotes the negative log-likelihood
that describes the imaging system physics and noise statistics.
The benefits of modeling the system noise and physics properties
were the primary driver
for the early work on MBIR methods for PET and SPECT,
compared to classical reconstruction methods like FBP
that use quite simple geometric models
and lack statistical modeling.
\addnew{In MRI,
early iterative methods were driven by non-Cartesian sampling
and parallel imaging
\cite{pruessmann:01:ais}.
}
The function
$p(\x)$
in \eqref{e,xh,map}
denotes a Bayesian prior
that captures assumptions
about the image \x.
Markov random field models
were particularly popular in early work;
these methods typically assign
higher prior probabilities
for images \x
where neighboring pixels
tend to have similar values
\cite{geman:84:srg,besag:86:ots},
\addnew{often using ``line sites''
to infer the presence of boundaries
between pixels
\cite{kao:98:ira},
sometimes with the guidance
of images from other modalities
of the same patient
(``anatomical priors'')
\cite{chen:91:sfi,gindi:93:bro}
\cite{cao:97:upk,liang:91:ags}.}

Although the term \emph{sparsity}
is uncommon
in papers about MRF models,
the ``older'' assumption
that neighboring pixels tend to have similar values
is quite closely related
to the ``newer'' assumption
that the differences between neighboring pixel values
tend to be sparse.

The form of \eqref{e,xh,map}
is equivalent%
\footnote{
\add{
For any prior $p(\x)$,
one can simply define
$\reg R(\x) = -\log p(\x)$
to write 
\eqref{e,xh,map}
in the form
\eqref{e,f+R}.
However,
for most regularizers $R(\x)$
used in practice,
defining
$p(\x) \propto \exp(- \reg \R(\x))$
would be an
``improper prior''
because there is no constant
that makes $p(\x)$ integrate to unity.
}}
to the following regularized optimization problem:
\begin{equation}
\xh = \argmin{\x} f(\x) + \reg R(\x)
\label{e,f+R}
,\end{equation}
where $f(\x)$ denotes a data-fidelity term
and $R(\x)$
denotes a regularizer
that encourages the image \xh
to have some assumed properties
such as piece-wise smoothness.
The positive regularization parameter \reg 
controls the trade-off between over-fitting the (noisy) data
and over-smoothing the image.
More recent MBIR papers,
and the commercial methods,
tend to adopt this regularization perspective
rather than using Bayesian terminology.
Early commercial PET and SPECT reconstruction methods
used unregularized algorithms
\cite{hudson:94:air},
but more recent methods
use edge-preserving regularization
involving
\addnew{nonquadratic functions of the}
differences between neighboring pixels
\cite{ahn:15:qco},
essentially implicitly assuming
that the image gradients are sparse,
\add{i.e., that those differences are mostly zero or near zero.}
In 1D, a typical regularizer would be
\begin{equation}
R(\x) = \textstyle 
\sum_{n=2}^N \psi(x_n - x_{n-1})
\label{e,Rx}
,\end{equation}
where $N$ is the number of pixels, and $\psi$ denotes a ``potential function''
(in Bayesian parlance)
such as the hyperbola
\(
\psi(z) = \sqrt{|z|^2 + \delta^2}
\)
\add{or a generalized Gaussian function
\cite{bouman:93:agg}.}
(A few modifications of the regularizer are needed
to make it work well in practice
\cite{fessler:96:srp,nuyts:02:acp}.)
MBIR methods for clinical CT systems
also use edge-preserving regularization
\cite{thibault:07:atd}.

The \add{potential functions $\psi$
that are used clinically
in FDA-approved methods
for CT and PET
include a generalized Gaussian function
\cite{thibault:07:atd}
and a relative-difference prior
\cite{nuyts:02:acp}.
These potential functions
}
are relatives
of the total variation (TV) regularizer
that is studied widely in the academic literature.
However, TV imposes a strong assumption of gradient sparsity
because it uses the nonsmooth absolute value potential
$\psi(z) = |z|$
that is well-suited to images that are piece-wise \emph{constant}
but less suitable for images that are piece-wise \emph{smooth}.
In particular,
the TV regularizer leads to CT images 
with undesirable patchy textures;
so the commercial systems
use an edge-preserving regularizer
that does not enforce sparsity as strictly
\cite{thibault:07:atd}.
In summary,
the current clinical methods for PET, SPECT and CT
use optimization formulations
of the form
\eqref{e,f+R}
with regularizers akin to
\eqref{e,Rx},
thereby moderately encouraging
gradient sparsity.

\section{Sparsity Using Mathematical Models}
\label{sec:sparsity}

This section discusses image reconstruction methods
that are based on models for the image \x
that involve some form of sparsity.
\add{Methods based on sparsity models
have a long history in signal processing,
e.g.,
\cite{leahy:91:otd,harikumar:96:ana}.
}
Such methods are now being used clinically
to accelerate MRI scans,
making such scans shorter,
reducing the effects of patient motion
and improving patient comfort.

The regularizer based on finite differences
in \eqref{e,Rx} \add{(e.g., with $\psi(z) = |z|$)}
is equivalent to assuming the image gradients are sparse.
\add{Assumptions of gradient sparsity
or piece-wise smoothness
have a long history in imaging
\cite{geman:84:srg,liang:89:hri,besag:86:ots}.
}
This model
is a special case
of the more general assumption
that $\T\x$ is sparse
for some spatial \add{operator} 
\T.
This is called ``\emph{analysis regularization}''
and a typical image reconstruction optimization formulation
for such models is
\begin{equation}
\xh = \argmin{\x} \frac{1}{2} \norm{\A \x - \y}_2^2 + \reg \normi{\T \x}
\label{e,Tx}
.\end{equation}
\add{The $\ell_1$ norm is often used as the sparsity regularizer, and can be viewed as a convex relaxation or convex envelope of the nonconvex $\ell_0$ ``norm" that measures the size of the support of a vector or counts the number of nonzero entries. Alternative penalties such as $\left \| \T \x \right \|_{p}^{p}$ for $0 < p < 1$ that better approximate $\left \| \T \x \right \|_{0}$ have also been used for reconstruction~\cite{chartrand09}.}
There are many \add{operators} \T that have been used for image reconstruction;
the two most popular ones are
finite-differences,
corresponding to TV,
and various wavelet transforms.
Wavelets are the \add{sparsifying} model
used in the JPEG 2000 image compression standard,
because they are effective
at sparsifying natural images.
The combination of both wavelets and TV
is particularly common in MRI
\cite{lustig2007sparse},
and,
although the details are proprietary,
it is likely that such combinations
are used in the commercial MRI systems,
\eg,~\cite{geertsossevoort:18:cs}.

\addnew{
In some settings one has a ``prior image'' $\bar{\x}$ available,
in which case one can modify
\eqref{e,Tx}
to encourage similarity with that prior image
using a cost function like:
\begin{equation}
\xh = \argmin{\x} \frac{1}{2} \norm{\A \x - \y}_2^2 + \reg \normi{\T (\x - \bar{\x})}
\label{e,piccs}
.\end{equation}
The prior image constrained compressed sensing (PICCS) approach
is an example of this type of approach
\cite{chen:08:pic}.
}

An alternative
to the analysis regularization 
model
\eqref{e,Tx}
is to assume that the image can be represented
as a sparse linear combination
of atoms from a dictionary,
\ie,
$ \x = \D \z $
where \D is a dictionary
and \z is a coefficient vector.
One way to express this assumption
as an optimization problem is
\begin{equation}
\xh = \D \zh
,\quad
\zh = \argmin{\z} \frac{1}{2} \normii{\A \D \z - \y}^2 + \reg \normi{\z}
\label{e,xh,zh}
,\end{equation}
where \A denotes the imaging system model.
\add{
This synthesis formulation
is equivalent to the analysis formulation
\eqref{e,Tx}
when \D and \T are both square and full-rank 
and
$\D = \T^{-1}$ (a basis).
When \T is tall and its rows form a frame, then \eqref{e,Tx} is equivalent to a synthesis formulation ($\D = \T^{L}$, a left-inverse of \T), but with \z in \eqref{e,xh,zh} constrained to be in the range space of \T.
But usually \T is a general operator and \D is wide.
}
A drawback of this synthesis sparsity formulation
is that it relies heavily on the assumption
that
$ \x = \D \z $,
whereas an approximate form
$ \x \approx \D \z $
may be 
more reasonable in practice,
particularly when the dictionary \D
comes from a mathematical model
that might not perfectly represent natural medical images.
An alternative synthesis formulation
that allows an approximate sparsity model is:
\[
\xh = \argmin{\x} \frac{1}{2} \norm{\A \x - \y}_2^2 + \reg R(\x)
,\]
\[
R(\x) = \min_{\z} \frac{1}{2} \norm{\x - \D \z}_2^2 + \alpha \norm{\z}_1
.\]
A 
drawback of this approach
is that it requires one to select two regularization parameters
($\alpha$ and \reg).
\add{Other 
sparsity models include generalized analysis models~\cite{sai2013tl},
and the balanced sparse model for tight frames \cite{liuqu15},
where the signal is sparse in a synthesis dictionary
and also approximately sparse in the corresponding transform 
(transpose of the dictionary) domain,
with a common sparse representation in both domains.
These models have been applied to inverse problems such as in compressed sensing MRI~\cite{lustig2007sparse,liuqu15,liuqu16}.}

The drawback of all of the models discussed
in this section
is that, traditionally, \add{the underlying operators such as}
\T and \D are designed mathematically,
\add{typically with empirical validation on real data},
rather than being
\add{computed directly from training data
or adapted to a specific patient's data}.
Nevertheless,
they are useful,
as evidenced by their adoption in clinical MRI systems.
\add{Most of the}
methods in subsequent sections
are 
more data-driven approaches.

\section{Low-rank Models}
\label{sec:lowrank}

While sparsity models have been popular in image reconstruction, particularly in CS, various alternative models exploiting properties such as the inherent low-rankness of the data have also shown promise in imaging applications.
This section reviews some of the low-rank models and their extensions
such as when combined with sparsity,
followed by recent structured low-rank matrix approaches%
~\cite{haldar2014low,jin2016general,lee2016acceleration,ongie2016off,ongie2017fast,jin2017mri}.
\addnew{The assumption that a matrix
is low-rank
is equivalent to assuming
that its singular values are sparse.
Both sparsity and low-rankness involve
assumptions of simplicity,
and these relationships are unified
by the notion of atomic norms
\cite{chandrasekaran:12:tcg}.
}
\add{As there are many works on low-rank models for various modalities and applications, we only provide an overview of some low-rank methods, their extensions, and connections to sparsity, rather than an exhaustive review.}

\subsection{Low-Rank Models and Extensions} \label{sec:lowrankpart1}

Low-rank models have been exploited in many imaging applications such as \add{dynamic MRI~\cite{zliang07}, functional MRI~\cite{singh15,chiew2016}, diffusion-weighted MRI~\cite{yuxinhu19}, and MR fingerprinting (MRF)~\cite{mazoreldar18,zhao18,cruz19}.}

\add{Low-rank} assumptions are especially useful when processing dynamic or time-series data, and have been popular in dynamic MRI, where the underlying image sequence tends to be quite correlated over time.
In dynamic MRI, the measurements are inherently undersampled because the object changes as the samples are collected.
Reconstruction methods therefore typically pool the k-t space data in time to make sets of k-space data (the underlying dynamic object is written in the form of a Casorati matrix~\cite{zliang07}, whose rows represent voxels and columns denote temporal \emph{frames}, and the sets of k-space data denote measurements of such frames) that appear to have sufficient samples.
However, these methods can have poor temporal resolution and artifacts due to pooling.
Careful model-based (CS-type) techniques can help achieve improved temporal or spatial resolution in such undersampled settings.

Several works have exploited low-rankness of the underlying Casorati (space-time) matrix for dynamic MRI reconstruction~\cite{zliang07,hald10,bzhao10,peder09}.
Low-rank modeling of local space-time image patches has also been investigated in~\cite{Trzaskolocallow11}.
Later works combined low-rank (L) and sparsity (S) models for improved reconstruction. Some of these works model the dynamic image sequence as both low-rank and sparse (L \& S)~\cite{lingala11,bzhao12}.
There has also been growing interest in models that decompose the dynamic image sequence into the sum of a low-rank and sparse component (a.k.a. robust principal component analysis (RPCA))~\cite{Can11,guo14}. In this L+S model, the low-rank component can capture the background or slowly changing parts of the dynamic object, whereas the sparse component can capture the dynamics in the foreground such as local motion or contrast changes, etc.

Recent works have applied the L+S model to dynamic MRI reconstruction~\cite{otazo15,ben14}, with the S component modeled as sparse by itself or in a known transform domain.
Accurate reconstructions can be obtained~\cite{otazo15} when the underlying L and S components are incoherent (distinguishable) and the k-t space acquisition is appropriately incoherent with these components.
The L+S reconstruction problem can be formulated as follows:
\begin{align}
\nonumber \min_{\x_{L},\, \x_{S}} &  \frac{1}{2} \left \| \A(\x_{L}+ \x_{S})-\y \right \|_{2}^{2} + \lambda_{L}\left \| \R_{1}(\x_{L}) \right \|_{*} \\
 &\; \; \; \; + \lambda_{S} \left \| \T \x_{S} \right \|_{1}. \label{L+S}
\end{align}
Here, the underlying vectorized object satisfies the L+S decomposition $\x = \x_{L} + \x_{S}$.
The sensing operator $\A$ acting on it can take various forms. For example, in parallel imaging of a dynamic object, $\A$ performs frame-by-frame multiplication by coil sensitivities (in the SENSE approach) followed by undersampled Fourier encoding.
The low-rank regularization penalizes the nuclear norm of $\R_{1}(\x_{L})$, where $\R_{1}(\cdot)$ reshapes its input into a space-time matrix. The nuclear norm serves as a convex surrogate or envelope for the nonconvex matrix rank.
The sparsity penalty on $\x_{S}$ has a similar form as in CS approaches, and $\lambda_{L}$ and $\lambda_{S}$ are non-negative weights above.
Problem~\eqref{L+S} is convex and can be solved using various iterative techniques. Otazo et al.~\cite{otazo15} used the proximal gradient method, wherein the updates involved simple singular value thresholding (SVT) for the L component and soft thresholding for the S component.
Later, we mention a data-driven version of the L+S model in Section~\ref{sec:datadriven:dictionarylearning}.
While the above works used low-rank models of matrices (e.g., obtained by reshaping the underlying multi-dimensional dynamic object into a space-time matrix), \add{other} recent works also used low-rank \add{tensor models of the underlying object (a tensor) in reconstruction~\cite{banco16,yaman17,he16,Christodoulou18}.}

\subsection{Low-Rank Structured  Matrix Models}
\label{sec:hankel}

\add{
So far, we discussed low-rank and sparsity models
that both 
involve dimensionality reduction.
The former involves a low-dimensional subspace,
whereas the latter is typically viewed as unions of such subspaces.
This section reviews the structured low-rank methods
and elaborates on 
the connections between sparsity and low-rank modeling.
}
 
\add{The low-rank Hankel structure matrix approaches \cite{jin2016general,lee2016acceleration,ongie2016off,lee2016reference,ongie2017fast,jin2017mri,ye2016compressive,ongie2017convex,haldar2014low}
have been studied extensively
for various imaging problems.
Unlike the standard low-rank approaches that are based on the redundancies between similar data, 
the low-rank Hankel structure matrix approaches
are based on the fundamental duality
between spatial domain sparsity
and the spectral domain Hankel matrix rank, which is 
also related to local polynomial approximation\cite{liang:89:hri}.}

\begin{figure}[!bt]
	\center
	{\includegraphics[width =0.5\textwidth]{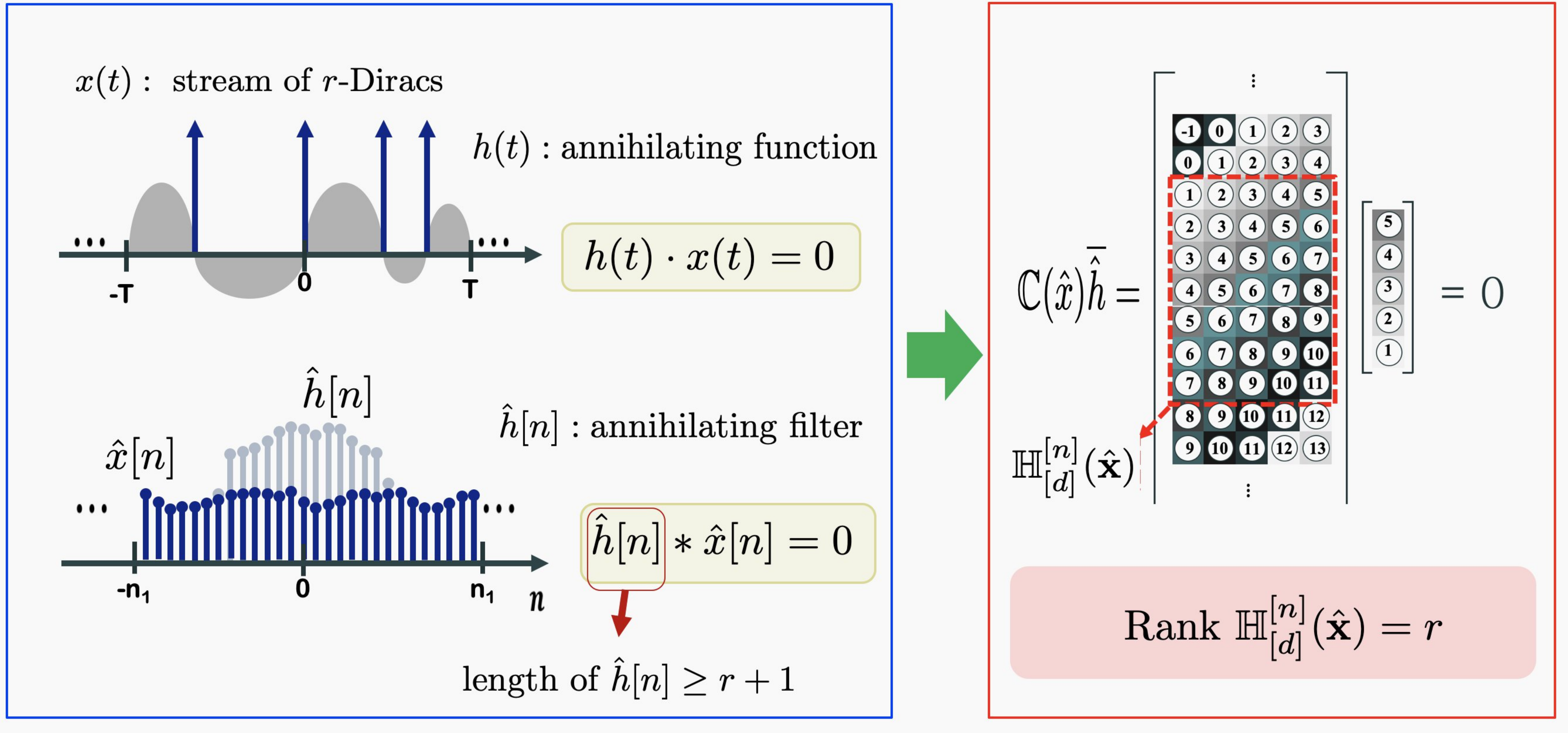}}~~
	\caption{\add{Fundamental duality between sparsity in the image domain and low-rank Hankel matrix in Fourier domain.}}
	\label{fig:dualty}
	\vspace{-0.15in}
\end{figure}

To explain this \add{duality},
we \add{first review} the literature
on the sampling theory of signals
having finite rate of innovations (FRI)
\cite{vetterli2002sampling,maravic2005sampling,liangslr}. 
Consider
a superposition of $r$ Dirac impulses as shown in Fig.~\ref{fig:dualty}:
\begin{equation}\label{eq:signal3}
x(t) = \sum_{j=0}^{r-1} a_j \delta \left( t- t_j \right),
\, \quad t_j \in [0, 1].
\end{equation}
The associated
Fourier series coefficients
are given by
\begin{eqnarray}\label{eq:fs}
\hat x[k] = \sum_{j=0}^{r-1} a_j e^{-i2\pi k t_j }
\ .
\end{eqnarray}
The sampling theory for FRI signals
\cite{vetterli2002sampling,maravic2005sampling}
showed that
there exists an
annihilating filter
$\hat h[k]$ in the Fourier domain,
of length $(r+1)$,
such that
\begin{eqnarray}\label{eq:annihilating}
(\hat h \ast \hat x)[k] = \sum_{l=0}^r \hat h[l]\hat x[k-l] = 0
,\end{eqnarray}
whose $z$-transform representation is given by
\begin{eqnarray}\label{eq:afilter}
\hat h(z) = \sum_{l=0}^r \hat h[l] z^{-l} = \prod_{j=0}^{r-1} (1- e^{-i2\pi t_j} z^{-1})
\ .
\end{eqnarray}

\add{As shown in Fig.~\ref{fig:dualty}, the annihilating filter relationship implies that the convolution matrix multiplied by an annihilating filter vector vanishes.
Accordingly, 
the following Hankel structured matrix,
corresponding to the submatrix of the convolution matrix,
is rank-deficient:}
    \begin{eqnarray*}
\hank_{[d]}^{[n]}(\hat  \x) :=\left[
        \begin{array}{cccc}
      \hat  x[0]  & \hat  x[1] & \cdots   & \hat  x[d-1]   \\
     \hat  x[1]  & \hat  x[2] & \cdots &   \hat  x[d] \\
         \vdots    & \vdots     &  \ddots    & \vdots    \\
      \hat  x[n-d]  & \hat  x[n-d+1] & \cdots & \hat  x[n-1]\\
        \end{array}
    \right] 
    \end{eqnarray*}
where $[n]:=\{0,\cdots, n-1\}$.
Specifically,
it was shown in \cite{ye2016compressive}
that if the minimum annihilating filter length is $r+1$,
then
$$
\rank \hank_{[d]}^{[n]}(\hat\x) = r.
$$
Thus,
given sparsely sampled spectral measurements
on the index set $\Omega \subset [n]$,
the missing spectrum estimation problem can be formulated as
\begin{eqnarray}\label{eq:slr}
\argmin{\mb\in \Cd^n}  &\| \hank_{[d]}^{[n]}(\mb)\|_* \\
\mbox{subject to } & P_\Omega(\mb) = P_\Omega(\hat  \x),
\end{eqnarray}
where  $P_\Omega(\cdot)$ denotes the projection on the measured
k-space samples on the index set $\Omega$.
Although the above discussion is for  Dirac impulses, the same principle holds for general FRI signals that can be converted to
Diracs or differentiated Diracs after a whitening operator, since the corresponding Fourier spectrum is a simple element-wise multiplication with the spectrums of the operator and the unknown signal,
and the weighted spectrum has a low-rank Hankel structure  \cite{jin2016general,ye2016compressive}.

In contrast to 
standard compressed sensing approaches
for MRI,
the optimization problem in \eqref{eq:slr}
is purely in the measurement domain. 
After estimating the fully sampled Fourier data, 
the final reconstruction
can be obtained by a simple inverse Fourier transform.
This property leads to remarkable flexibility in real-world applications
\add{that} classical approaches have difficulty exploiting. 
For example,
this formulation has been successfully applied
to compressed sensing MRI with state-of-the art performance for single coil imaging
\cite{haldar2014low,jin2016general,lee2016acceleration,ye2016compressive,ongie2016off}.  
Another flexibility is the recovery of images
from multichannel measurements with unknown sensitivities
\cite{shin2014calibrationless,jin2016general}.
These schemes rely on the low-rank structure of a structured matrix,
obtained by concatenating block Hankel matrices
formed from each channel's data. 
Similar to L+S decomposition in~\cite{otazo15},
the L+S model for Hankel structure matrix
was also used to remove the $k$-space outliers in MR imaging problems~\cite{jin2017mri}.
\add{
Such approaches have also been successfully used
for super-resolution microscopy~\cite{min2018grid},
image inpainting problems~\cite{jin2015annihilating},
image impulse noise removal~\cite{jin2018sparse}, etc.}

The common thread between
\add{sparsity models, low-rank, and structured low-rank}
models
is that they all
strive to capture signal redundancies
to make up for missing or noisy data.

\section{Data-Driven and Learning-Based Models}
\label{sec:datadriven}

The most recent class of methods
constituting the fourth \add{category of} image reconstruction \add{schemes}
exploit data-driven and learning-based models.
This section and the next
review \add{several of} these varied models and methods. \add{The development of efficient algorithms for the often nonconvex learning-based problems is briefly discussed.}
The pros and cons of various methods, as well as their connections are also discussed.

\subsection{Partially Data-Adaptive Sparsity-Based Methods}

While early reconstruction methods such as in CS MRI 
used sparsity in known transform domains
such as wavelets~\cite{lustig2007sparse},
total variation domain,
contourlets~\cite{qu2010iterative}, etc.,
later works proposed partially data-adaptive sparsity models by incorporating directional information of patches or block matching, etc., during reconstruction.

\addnew{A patch-based directional wavelets (PBDW) scheme was proposed for MRI in~\cite{qu2012undersampled},}
wherein the regularizer was based on analysis sparsity and was the sum of the $\ell_1$ norms
of each optimally (adaptively) rearranged and transformed (by fixed 1D Haar wavelets) image patch.
The patch rearrangement or permutation involved rearranging pixels parallel to a certain geometric direction, approximating patch rotation.
The best permutation for each patch
from among a set of pre-defined permutations
was pre-computed based on \addnew{initial reconstructions}
to minimize the residual between the transformed permuted patch and its \addnew{thresholded
version.}
\addnew{An improved 
reconstruction method was proposed in~\cite{ning2013magnetic},}
where the optimal permutations were computed for patches extracted
from the subbands in the 2D Wavelet domain
\addnew{(a shift-invariant discrete wavelet transform is used)} of the image.
\addnew{A recent work~\cite{zhan2016fast} proposed a different \add{effective modification} of the PBDW scheme,}
wherein a unitary matrix is 
\addnew{adapted to sparsify the patches
grouped with a common (optimal) permutation.
In this case, the analysis sparsity penalty during reconstruction used the $\ell_1$ norms of patches
transformed by the adapted unitary matrices (one per group of patches).}

A different \add{fast and effective 
\addnew{method} (\addnew{patch-based} nonlocal operator or PANO)} was proposed in~\cite{qu2014magnetic},
wherein for each patch, a small group of patches most similar to it was pre-estimated
(called \emph{block matching}),
and the regularizer during reconstruction
penalized the sparsity of the groups of 
\addnew{patches in a known transform domain.}
\add{Another reconstruction scheme based on adaptive clustering of patches
was applied to MRI in~\cite{akcakaya:11:lds}.}
All these 
aforementioned methods
\add{are also} \addnew{quite} related to the recent transform learning-based methods
described in Section~\ref{sec:datadriven:transformlearning},
where the sparsifying operators are fully adapted in an optimization framework.
\vspace{-0.1in}
\subsection{Synthesis Dictionary Learning-Based Approaches for Reconstruction} \label{sec:datadriven:dictionarylearning}


Among the learning-based approaches that have shown promise for medical image reconstruction, one popular class of methods exploits \emph{synthesis dictionary learning}.

\subsubsection{Synthesis Dictionary Model}

As briefly discussed in Section~\ref{sec:sparsity},
the synthesis model suggests that a signal can be approximated
by a sparse linear combination of atoms or columns of a dictionary,
i.e., the signal lives approximately in a subspace spanned by a few dictionary atoms.
Because different signals may be approximated with different subsets of dictionary columns,
the model is viewed as a union of subspaces model~\cite{vidal11}.

\begin{figure}[!t]
\hspace{-0.0in}\includegraphics[width=0.47\textwidth]{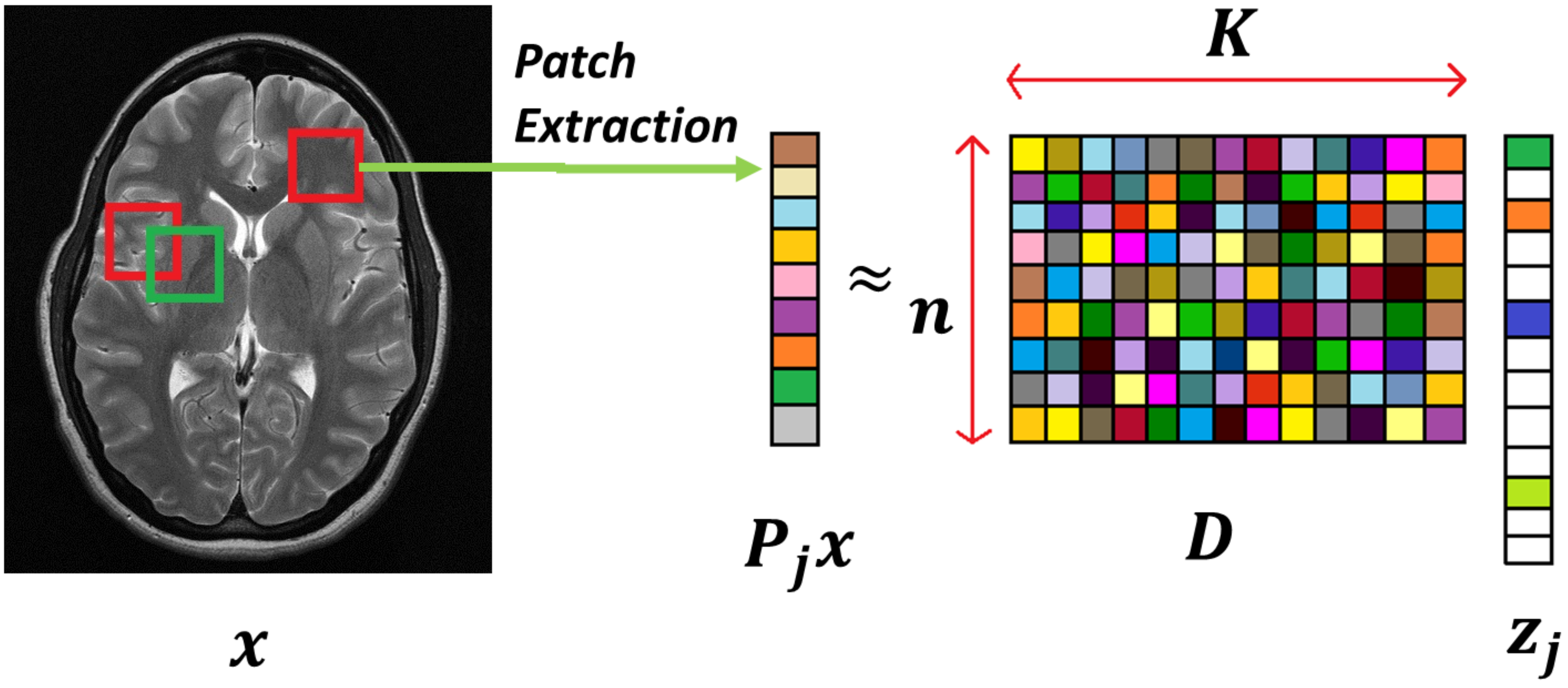}
\caption{The synthesis dictionary model for image patches: overlapping patches $\P_j \x$ of the image $\x$ are assumed approximated by sparse linear combinations of columns of the dictionary $\D$, i.e., $\P_j \x \approx \D \mathbf{z}_j$, where $\mathbf{z}_j$ has several zeros (denoted with white blocks above). }
\label{fig:synthesisdictionarymodel}
\end{figure}

In imaging, the synthesis model is often applied to image patches (see Fig.~\ref{fig:synthesisdictionarymodel}) or image blocks $\P_j \x$ as $\P_j \x \approx \D \mathbf{z}_j$, with $\P_j$ denoting the operator that extracts a vectorized patch (with $n$ pixels) of $\x$, $\D \in \mathbb{C}^{n \times K}$ denoting a synthesis dictionary (in general complex-valued), and $\mathbf{z}_j \in \mathbb{C}^{K}$ being the sparse representation or code for the patch $\P_j \x$ with many zeros.
While dictionaries based on the discrete cosine transform (DCT), etc., can be used to model image patches, much better representations can be obtained by adapting the dictionaries to data.
The learning of synthesis dictionaries has been explored in many works~\cite{ols96,Aharon06,Mai10} and shown to be promising in inverse problem settings~\cite{elad06,Mai08,sai2011dlmri}.

\subsubsection{Dictionary Learning for MRI}

\addnew{A dictionary learning-based method for MRI (DL-MRI) was proposed in~\cite{sai2011dlmri},}
where the image and the dictionary for its patches are simultaneously estimated
from limited measurements.
The approach also known as blind compressed sensing (BCS)~\cite{eldar2011bcs}
does not require training data
and learns a dictionary that is highly adaptive to the underlying image content.
However, the optimization problem is highly nonconvex, and is formulated as follows:
\begin{align}
\nonumber & \min_{\x,\D,\Z}  \frac{1}{2} \left \| \A \x - \y \right \|_{2}^{2} + \beta \sum_{j=1}^{N}\left \| \P_{j} \x - \D \mathbf{z}_{j} \right \|_{2}^{2} \\
 & \;\;\;\; \text{s.t.} \;\; \begin{Vmatrix}
\mathbf{z}_j
\end{Vmatrix}_{0} \leq s, \; \begin{Vmatrix}
\mathbf{d}_i
\end{Vmatrix}_{2} = 1, \; \forall \, i,j. \label{dlmri}
\end{align}
This corresponds to using a dictionary learning regularizer (weighted by $\beta>0$) of the following form:
\begin{align}
\nonumber  R(\x) = & \min_{\D,\Z} \sum_{j=1}^{N}\left \| \P_{j} \x - \D \mathbf{z}_{j} \right \|_{2}^{2} \\
& \;\; \text{s.t.} \;\; \begin{Vmatrix}
\mathbf{z}_j
\end{Vmatrix}_{0} \leq s, \; \begin{Vmatrix}
\mathbf{d}_i
\end{Vmatrix}_{2} = 1, \; \forall \, i,j, \label{dlmrireg}
\end{align}
where $\Z$ is a matrix whose columns are the sparse codes $\mathbf{z}_j$ that each have at most $s$ non-zeros, and the $\ell_0$ ``norm" counts the total number of nonzeros in a vector or matrix.
The columns $\mathbf{d}_i$ of $\D$ are constrained to have unit norm as otherwise $\mathbf{d}_i$ can be scaled arbitrarily along with corresponding inverse scaling of the $i$th row of $\Z$, and the objective is invariant to this \emph{scaling ambiguity}.

Problem~\eqref{dlmri} was optimized in~\cite{sai2011dlmri} by alternating between solving for the image $\x$ (image update step) and optimizing the dictionary and sparse coefficients (dictionary learning step).
\add{In} specific cases such as in single coil \add{Cartesian MRI,} 
\add{the image update step is solved} in closed-form using FFTs. 
However, the dictionary learning \add{step} 
\add{involves} a nonconvex and NP-hard 
\add{optimization} problem~\cite{ambruck09}.
Various dictionary learning algorithms exist for this problem and its variants~\cite{Aharon06,Rubinstein10,Mai10} that \add{often} alternate 
\add{between updating the sparse coefficients (sparse coding)} 
\add{and the dictionary.}
\add{The DL-MRI method for~\eqref{dlmri} used the K-SVD dictionary learning algorithm~\cite{Aharon06}} 
\add{and} showed significant image quality improvements
over previous CS MRI methods that used nonadaptive wavelets and total variation~\cite{lustig2007sparse}.
However, it is slow due to expensive and repeated sparse coding steps,
and lacked convergence guarantees.
In practice, variable rather than common sparsity levels across patches
can be allowed in DL-MRI
by using an error threshold based stopping criterion when \add{sparse coding with OMP~\cite{pati93}.}


\subsubsection{Other Applications and Variations}

Later works applied dictionary learning to dynamic MRI~\cite{lingaljacob13,wangying14,josecab14}, parallel MRI~\cite{weller16}, and PET reconstruction~\cite{Chen2015}.
An alternative Bayesian nonparametric dictionary learning approach
was used for MRI reconstruction in~\cite{huang14}.
\addnew{Dictionary learning was studied for CT image reconstruction in~\cite{xu:12:ldx}, which compared the BCS approach to pre-learning} the dictionary from a dataset
and fixing it during reconstruction.
The former was found to be more promising when sufficient views (in sparse-view CT) were measured,
whereas with very few views
(or with very little measured information),
pre-learning performed better.
Tensor-structured (patch-based) dictionary learning has also been exploited recently
for dynamic CT~\cite{Tan2015} and spectral CT~\cite{zhang17} reconstructions.

\subsubsection{Recent Efficient Dictionary Learning-Based Methods}

Recent work proposed efficient dictionary learning-based reconstruction algorithms, dubbed \add{SOUP-DIL image reconstruction algorithms}~\cite{ravrajfes17} that used the following regularizer:
\begin{align}
  & \min_{\D,\Z} \sum_{j=1}^{N}
\begin{Bmatrix}
\left \| \P_{j} \x - \D \mathbf{z}_{j} \right \|_{2}^{2} + \lambda^2 \begin{Vmatrix}
\mathbf{z}_j
\end{Vmatrix}_{0}
\end{Bmatrix}
\;\; \text{s.t.} \;\; \begin{Vmatrix}
\mathbf{d}_i
\end{Vmatrix}_{2} = 1, \; \forall \, i. \label{dlmrireg2}
\end{align}
Here, the \emph{aggregate} sparsity penalty $\sum_{j=1}^{N} \left \| \mathbf{z}_{j} \right \|_{0}$ 
with weight $\lambda^2$ 
automatically enables variable sparsity levels across patches.
The dictionary learning step of \add{the SOUP-DIL reconstruction algorithm efficiently} optimized \eqref{dlmrireg2} using an \emph{exact} block coordinate descent scheme by decomposing $\D \Z$ as a sum of outer products (SOUP) of dictionary columns and rows of $\Z$, and solving for $\mathbf{d}_i$ \add{and then the $i$th row of $\Z$ (by thresholding) in closed-form, and cycling over all such pairs ($1 \leq i \leq K$).}

\begin{figure}[!t]
\begin{center}
\begin{tabular}{cc}
\includegraphics[height=1.5in]{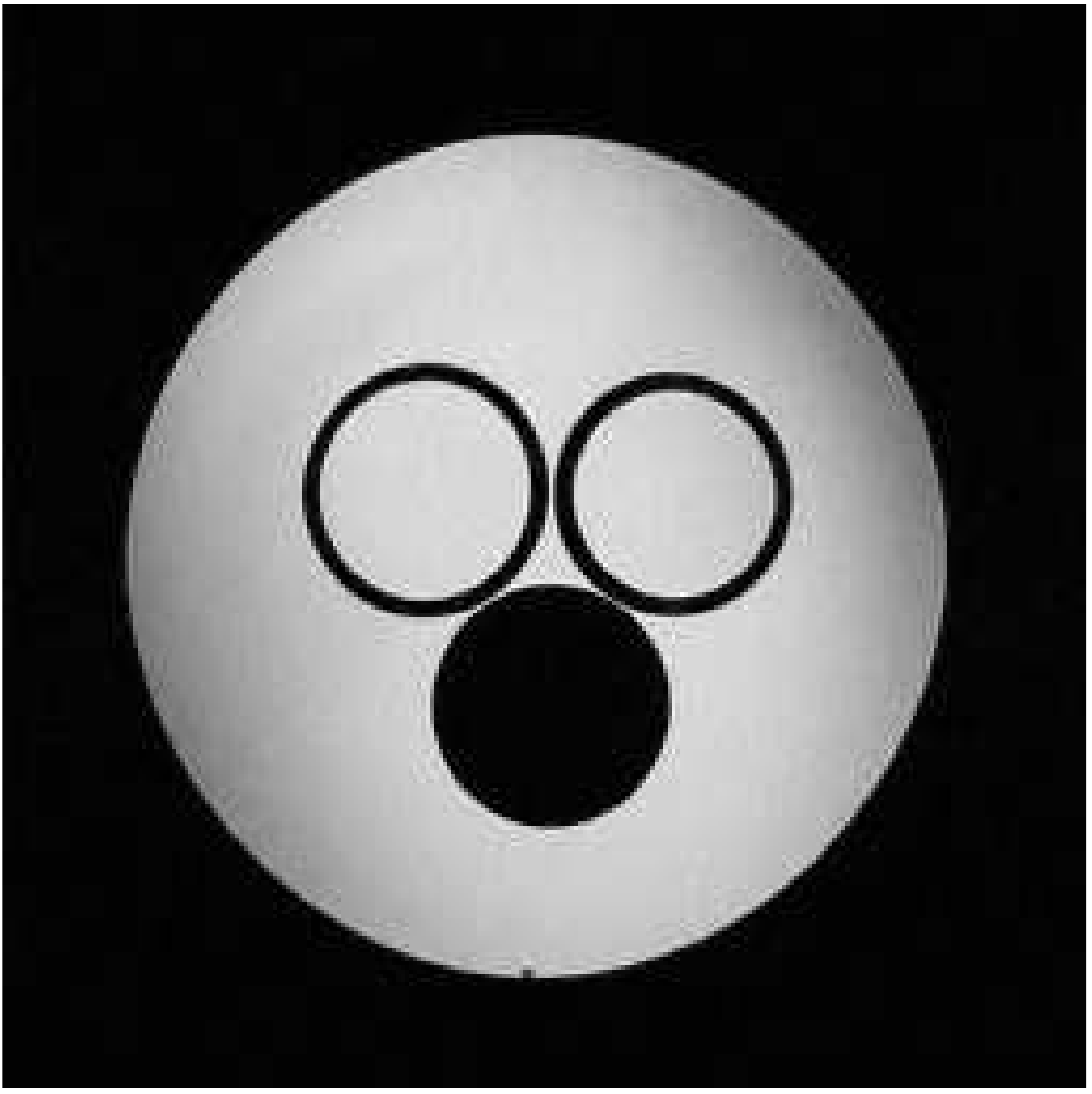}&
\includegraphics[height=1.5in]{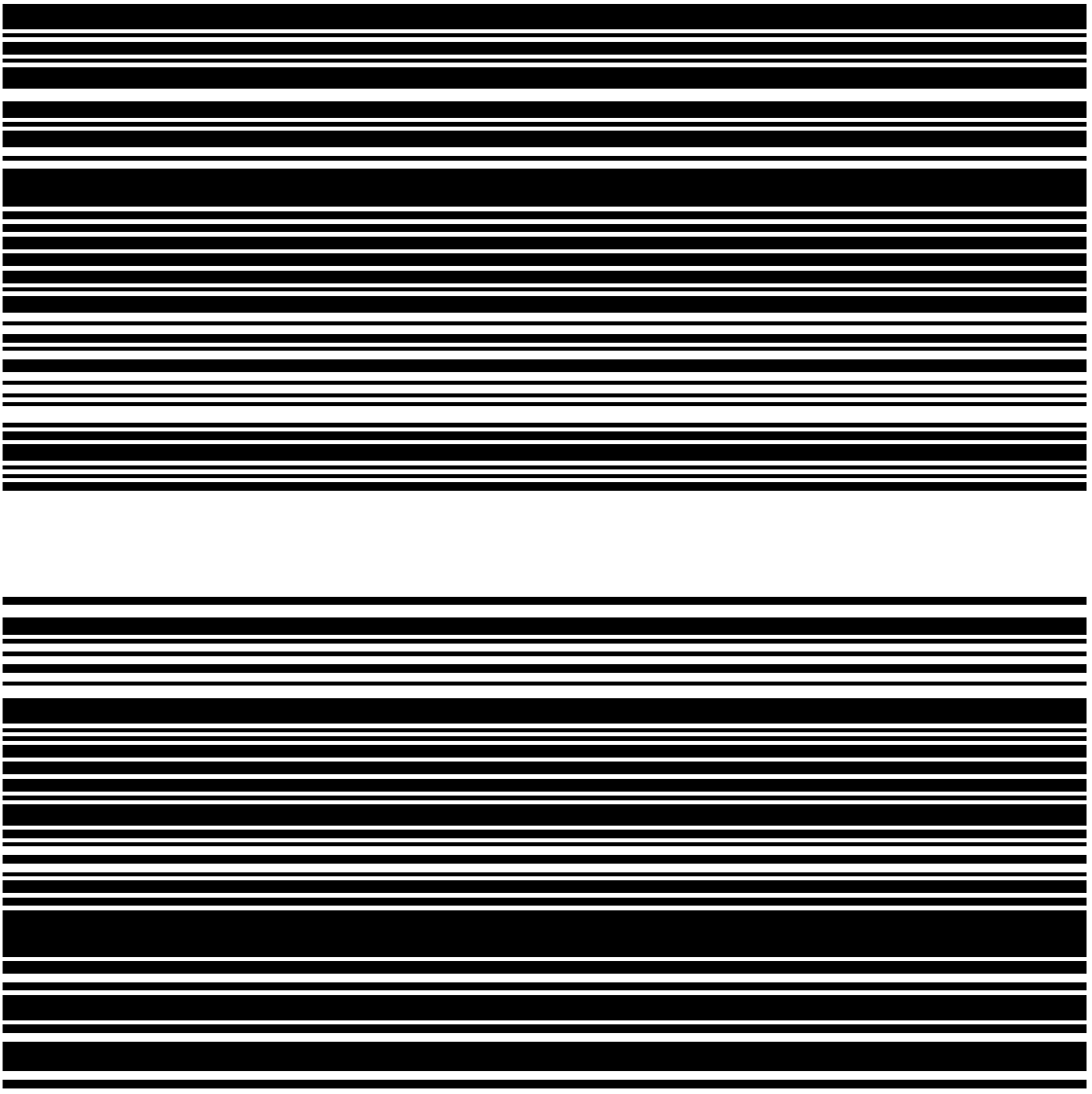}\\
(a) & (b) \\
\includegraphics[height=1.5in]{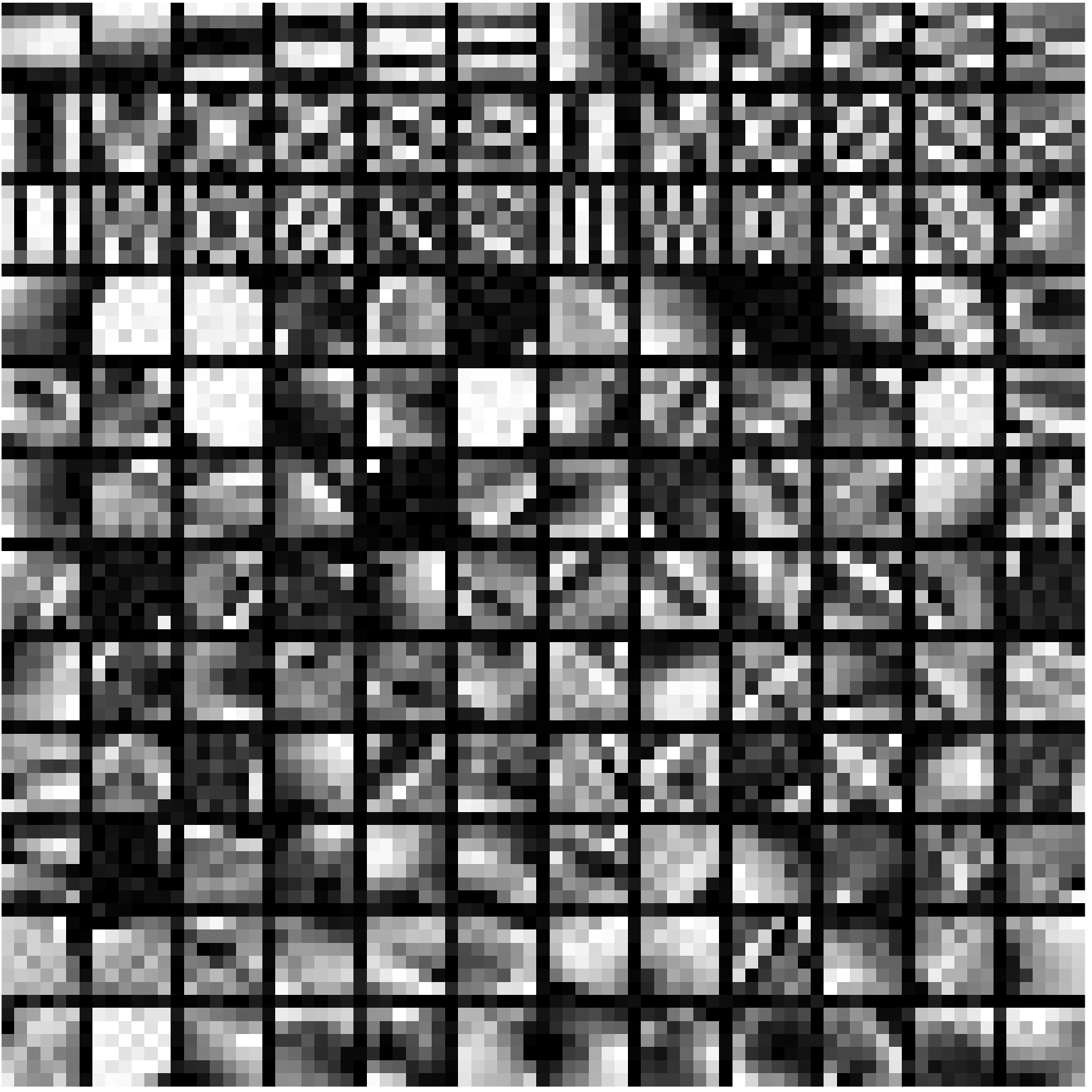}&
\includegraphics[height=1.5in]{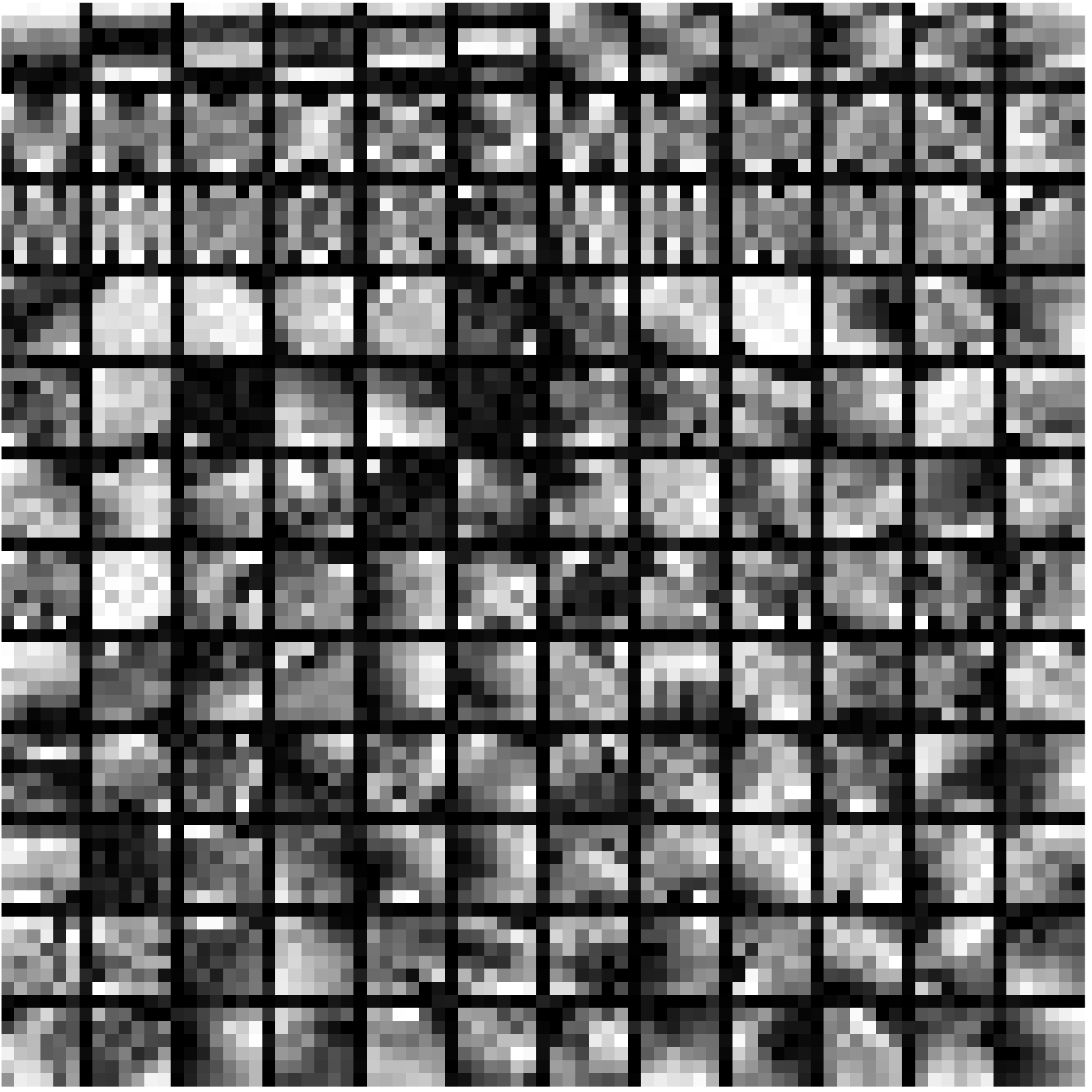}\\
(c) & (d)\\
\end{tabular}
\caption{Dictionary Learning for MRI (images from~\cite{ravrajfes17}): (a) SOUP-DILLO MRI~\cite{ravrajfes17} reconstruction \add{(with $\ell_0$ penalty)} of the water phantom~\cite{ning2013magnetic}; (b) sampling mask in k-space with 2.5x undersampling; and (c) real and (d) imaginary parts of the dictionary learned during reconstruction, with atoms shown as $6 \times 6$ patches.}
\label{fig:soupdilmriexample}
\end{center}
\end{figure}

While the earlier DL-MRI used inexact (greedy) and expensive sparse code updates and lacked convergence analysis, \add{the SOUP-DIL scheme} used efficient, exact updates and was proved to converge to the critical points (generalized stationary points) of the underlying problems and improved image quality over several 
schemes~\cite{ravrajfes17}.
Fig.~\ref{fig:soupdilmriexample} shows an example reconstruction with this BCS method along with the learned dictionaries.
Another recent work~\cite{ravmoorerajfes17} extended the L+S model for dynamic 
\add{image} 
reconstruction in \eqref{L+S} to a low-rank and adaptive sparse \addnew{signal 
model} that incorporated a dictionary learning regularizer similar to \eqref{dlmrireg2} for the $\x_{S}$ component.

\subsubsection{Alternative Convolutional Dictionary Model}

One can replace the patch-based dictionary model
with a convolutional model as
$\x \approx \sum_{i=1}^{K} \mathbf{d}_{i} \otimes \mathbf{c}_{i}$
that directly represents the image
as a sum of (possibly circular) convolutions of dictionary filters $\mathbf{d}_{i}$
and sparse coefficient maps $\mathbf{c}_{i}$%
~\cite{garcbrendt18,wohlberg16}.
The convolutional synthesis dictionary model is distinct from the patch-based model.
However, its main drawback is the inability to represent \addnew{very} low-frequency content in images,
necessitating 
pre-processing of images
to remove \addnew{very} low-frequency content prior to convolutional dictionary learning.
The utility of convolutional synthesis dictionary learning for biomedical image reconstruction
is an open and interesting area for future research;
see~\cite{chun:18:cdl}
for a denoising formulation
that could be extended to inverse problems.

\subsection{Sparsifying Transform Learning-Based Methods}
\label{sec:datadriven:transformlearning}

Several recent works have studied the learning of the 
efficient \emph{sparsifying transform model}
for biomedical image reconstruction~\cite{saibressiam15,sravTCI1,zheng:18:pua}.
This subsection reviews these advances (see~\cite{wensailukebres19} for an MRI focused review).

\subsubsection{Transform Model}

The sparsifying transform model is a generalization~\cite{sai2013tl} of the analysis dictionary model. The latter assumes that applying an operator $\W$ to a signal $\mathbf{f}$ produces several zeros in the output, i.e., the signal lies in the null space of a subset of rows of the operator.
The sparsifying transform model allows for a sparse approximation as $\W \mathbf{f} = \mathbf{z} + e$, where $\mathbf{z}$ has several zeros and $e$ is a 
\add{transform} domain modeling error.
Natural images are well-known to be approximately sparse in transform domains such as the DCT and wavelets, a property that has been exploited for image compression~\cite{marcellin00}, denoising, and inverse problems. 
A key advantage of the sparsifying transform model compared to the synthesis dictionary model is that the transform domain sparse approximation can be computed exactly and cheaply by thresholding $\W \mathbf{f}$~\cite{sai2013tl}.

\subsubsection{Early Efficient Transform Learning-Based Methods}

Recent works~\cite{saibressiam15,sravTCI1} proposed \addnew{transform learning based}
\add{image reconstruction}
methods that involved computationally cheap, closed-form updates in the iterative algorithms. 
\add{The} following square transform \add{learning~\cite{sai2013tl}} regularizer was used for reconstruction in~\cite{saibressiam15}:
\begin{align}  
R(\x) = & \min_{\W,\Z} \sum_{j=1}^{N}\left \| \W \P_{j} \x - \mathbf{z}_{j} \right \|_{2}^{2} + \gamma Q(\W) \;\; \text{s.t.} \;\; \begin{Vmatrix}
\Z
\end{Vmatrix}_{0} \leq s, \label{stlmrireg}
\end{align}
where $\W \in \mathbb{C}^{n \times n}$
is a square matrix and the transform learning regularizer
$Q(\W) = - \log \left | \det \W \right | + 0.5\left \| \W \right \|_{F}^{2}$ with weight $\gamma>0$
prevents trivial solutions in 
learning 
such as the zero matrix
or matrices with repeated rows.
Moreover, it also helps control the condition number of the transform~\cite{sai2013tl}\add{.}
The term
$\sum_{j=1}^{N}\left \| \W \P_{j} \x - \mathbf{z}_{j} \right \|_{2}^{2}$
denotes the transform domain modeling error or \emph{sparsification error},
which is minimized to learn a good sparsifying transform. %
The constraint in~\eqref{stlmrireg} on the $\ell_0$ ``norm" of the matrix $\Z$
controls the net or aggregate sparsity of all patches' sparse coefficients. 



The image reconstruction problem with regularizer~\eqref{stlmrireg}
was solved in~\cite{saibressiam15}
using a highly efficient block coordinate descent (BCD) approach
that alternates between \add{minimizing with respect to} 
$\Z$ (transform sparse coding step),
$\W$ (transform update step), and $\x$ (image update step).
Importantly, the transform sparse coding step has a closed-form solution,
where the matrix $\B$, whose columns are $\W \P_{j} \x$,
is thresholded to its $s$ largest magnitude elements,
with other entries set to zero.
\add{When the sparsity constraint is replaced with alternative sparsity promoting functions such as the $\ell_0$ sparsity penalty 
or $\ell_1$ penalty, the sparse coding solution is obtained in closed-form by hard 
or soft thresholding.}
\add{The transform update step} has a \add{simple solution} 
involving the singular value decomposition (SVD) of a small matrix~\cite{saibressiam15} \add{and} the image update step involves \add{a least squares problem (e.g., in the case of single coil Cartesian MRI, it is solved in closed-form using FFTs~\cite{saibressiam15}).}
This efficient BCD scheme was proven to converge in general to the critical points of the nonconvex reconstruction problem~\cite{saibressiam15}.


In practice, the sparsity controlling parameter can be varied over algorithm iterations
(a continuation strategy),
allowing for faster artifact removal initially
and then reduced bias over the iterations~\cite{sravTCI1}.
\add{The scheme in~\cite{saibressiam15} was shown}
to be much faster than the previous DL-MRI scheme.
Tanc and Eksioglu~\cite{tanc16} further combined \add{transform learning}
with global sparsity regularization
in known transform domains for CS MRI.

\addnew{Square transform learning has also been applied to CT reconstruction~\cite{lukebresler14}.}
Another recent work used \add{square transform learning} for low-dose CT image reconstruction~\cite{yeravyongfes17} with a shifted-Poisson likelihood penalty for the data-fidelity term in the cost (instead of the conventional weighted least squares penalty), but pre-learned the transform from a dataset and fixed it during reconstruction to save computation.



\addnew{Other works have explored alternative formulations for transform learning (e.g., 
overcomplete or tall~\cite{ravbres13} transforms)  that could be potentially used for image reconstruction.}

\subsubsection{Learning Rich Unions of Transforms for Reconstruction}

Since, images typically contain a diversity of textures, features, and edge information,
recent works~\cite{wensaibres15,sravTCI1,zheng:18:pua}
learned a union of transforms (a rich model) for image reconstruction.
In this setting, a collection of $K$ transforms are learned and the image patches are grouped or \emph{clustered} into $K$ classes, with each class of (\emph{similar}) patches best matched to and using a particular transform.
The \add{UNITE (UNIon of Transforms lEarning) image reconstruction} 
formulation in~\cite{sravTCI1} uses the following \add{regularizer:} 
\begin{align}  
\nonumber \hspace{-0.05in} R(\x) = & \min_{\begin{Bmatrix}
\W_{k},C_{k},\mathbf{z}_{j}
\end{Bmatrix}} \sum_{k=1}^{K} \sum_{j \in C_{k}}
\begin{Bmatrix}
\left \| \W_{k} \P_{j} \x - \mathbf{z}_{j} \right \|_{2}^{2} + \lambda^{2} \begin{Vmatrix}
\mathbf{z}_j
\end{Vmatrix}_{0}
\end{Bmatrix} \\
& \;\;\;\;\;\;\;\;\; \text{s.t.} \;\; \W_{k}^{H}\W_{k}= \mathbf{I} \; \forall \, k, \; \begin{Bmatrix}
C_{k}
\end{Bmatrix} \in G.
\label{unitemrireg}
\end{align}
Here, $C_{k}$ is a set containing the indices of all patches matched to the transform $\W_{k}$, and $G$ denotes the set of all partitions of $[1 : N ]$ into $K$ disjoint subsets, where $N$ is the total number of overlapping patches.
Note that when indexed variables are enclosed in braces (in \eqref{unitemrireg} and later equations), we mean the set of all variables over the range of the indices.


The \add{UNITE} 
reconstruction formulation jointly learns a collection of transforms, clusters and sparse codes patches, and reconstructs the image $\x$ from measurements.
\addnew{An} \add{efficient BCD algorithm with convergence
guarantees was proposed for optimizing the problem 
in~\cite{sravTCI1}.} 
\add{The $K$ transforms in~\eqref{unitemrireg} are} 
\add{unitary, which simplifies the BCD updates.}
\add{For MRI, UNITE-MRI achieved improved image quality over 
the square transform learning-based scheme}
when reconstructing from undersampled k-space measurements~\cite{sravTCI1}.


\begin{figure}[!t]
	\centering  	
    \begin{tikzpicture}
	[spy using outlines={rectangle,green,magnification=2,size=10mm, connect spies}]
	\node {\includegraphics[width=0.21\textwidth]{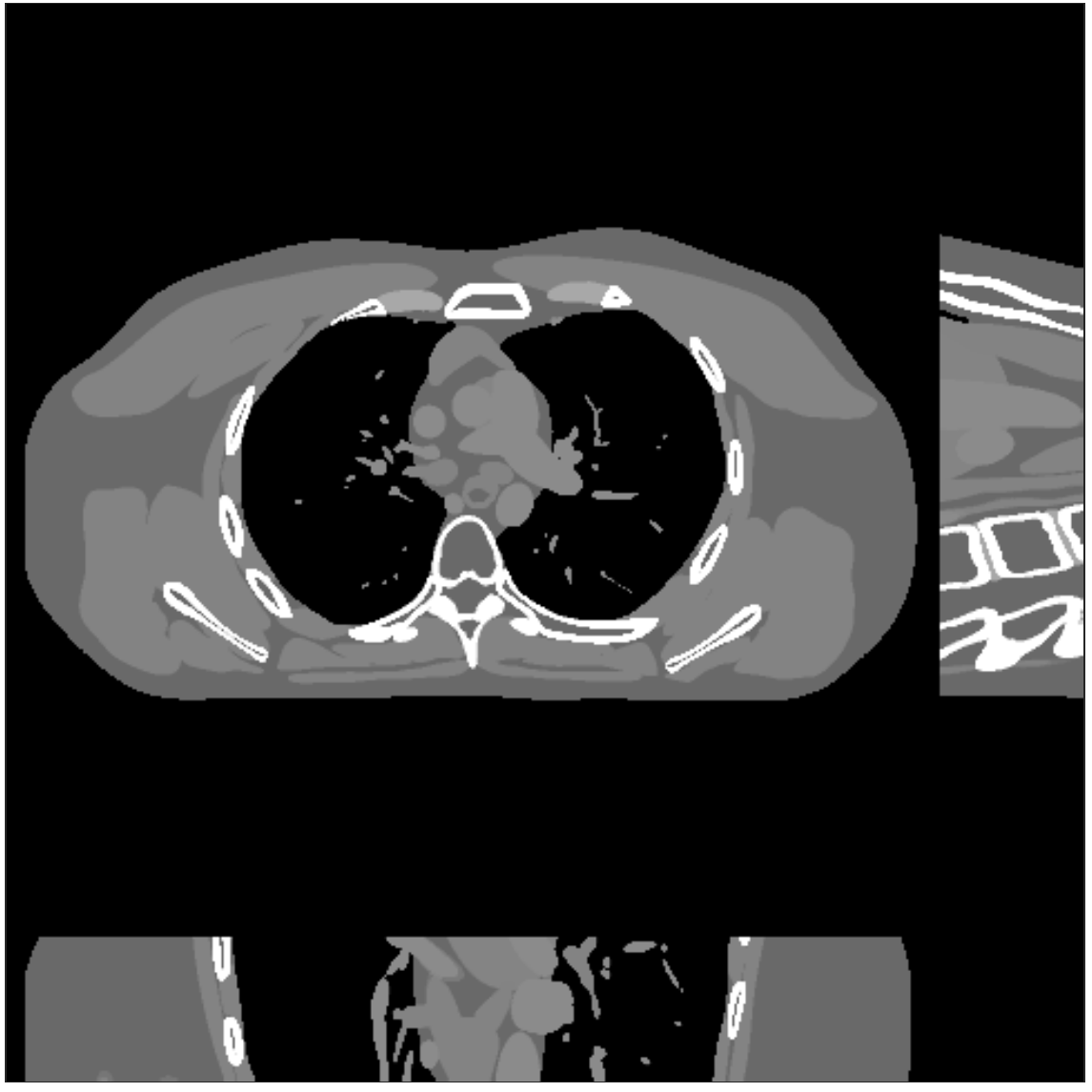}	};
	\spy on (0.87,0.12) in node [left] at (2.1,-1.2);	
	\spy on (-0.25,0.4) in node [left] at (2.1,1.65);		
	\end{tikzpicture}
	\begin{tikzpicture}
	[spy using outlines={rectangle,green,magnification=2,size=10mm, connect spies}]
	\node {\includegraphics[width=0.21\textwidth]{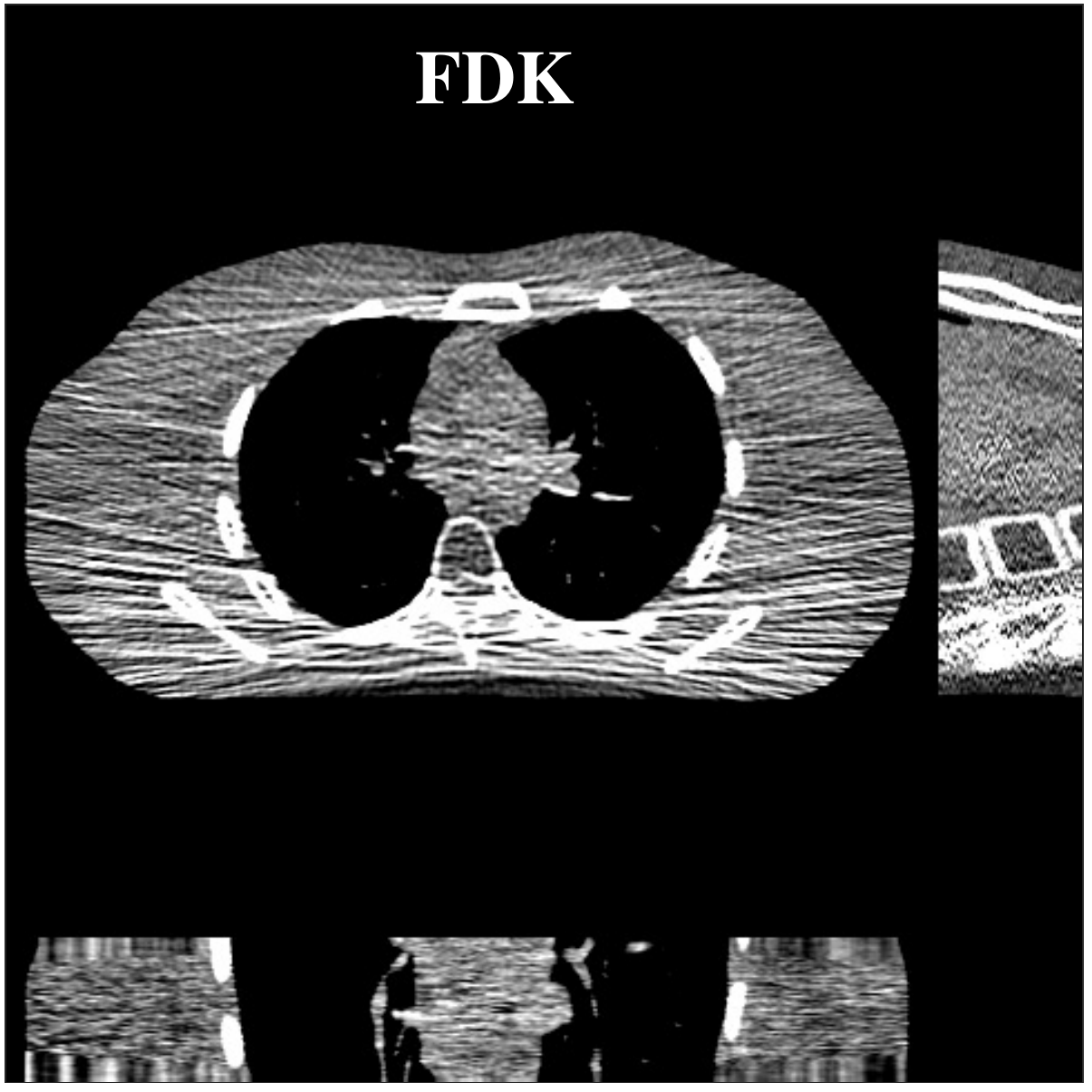}	};
	\spy on (0.87,0.12) in node [left] at (2.1,-1.2);	
	\spy on (-0.25,0.4) in node [left] at (2.1,1.65);		
	\end{tikzpicture}\\
	\begin{tikzpicture}
	[spy using outlines={rectangle,green,magnification=2,size=10mm, connect spies}]
	\node {\includegraphics[width=0.21\textwidth]{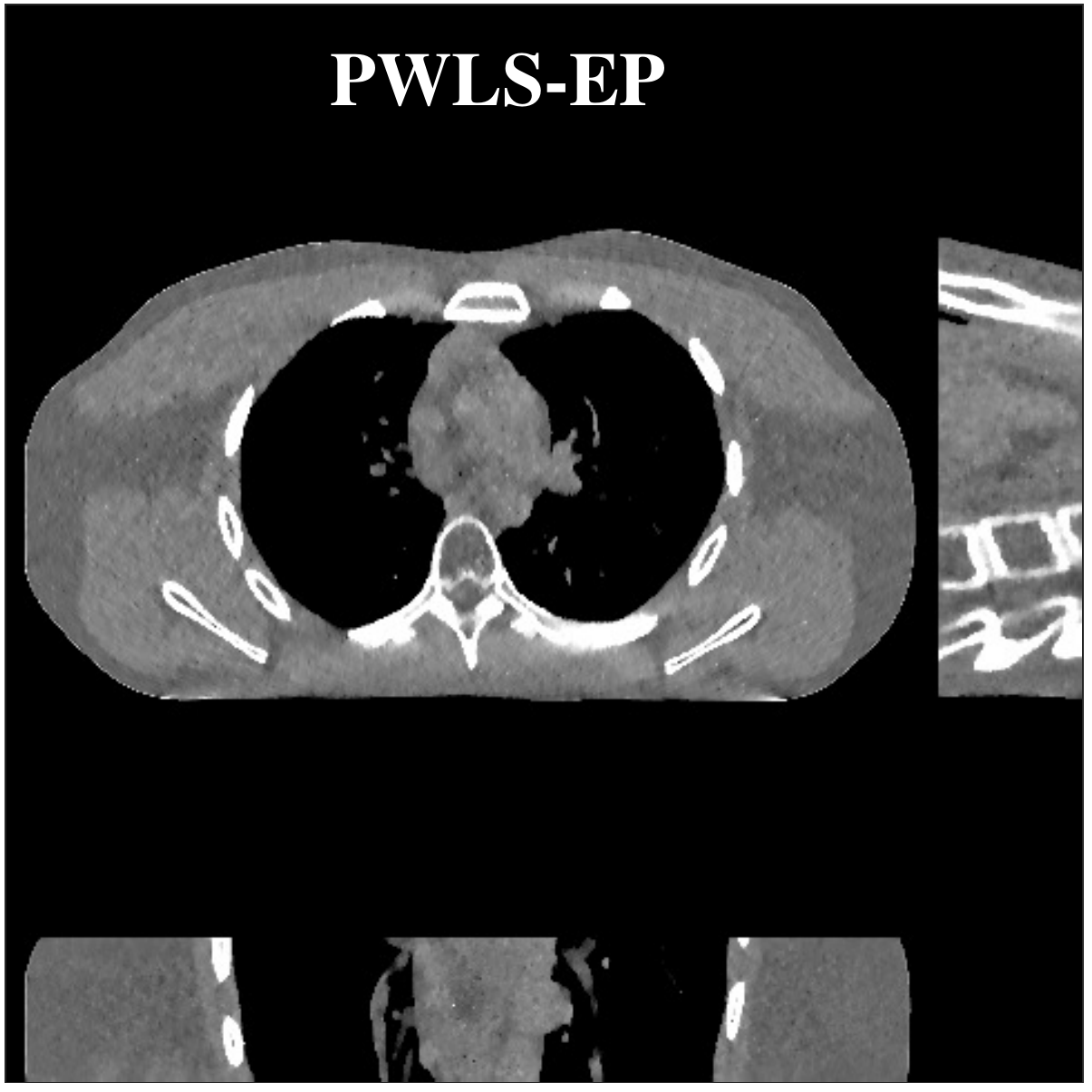}	};
	\spy on (0.87,0.12) in node [left] at (2.1,-1.2);		
	\spy on (-0.25,0.4) in node [left] at (2.1,1.65);		
	\end{tikzpicture}
	\begin{tikzpicture}
	[spy using outlines={rectangle,green,magnification=2,size=10mm, connect spies}]
	\node {\includegraphics[width=0.21\textwidth]{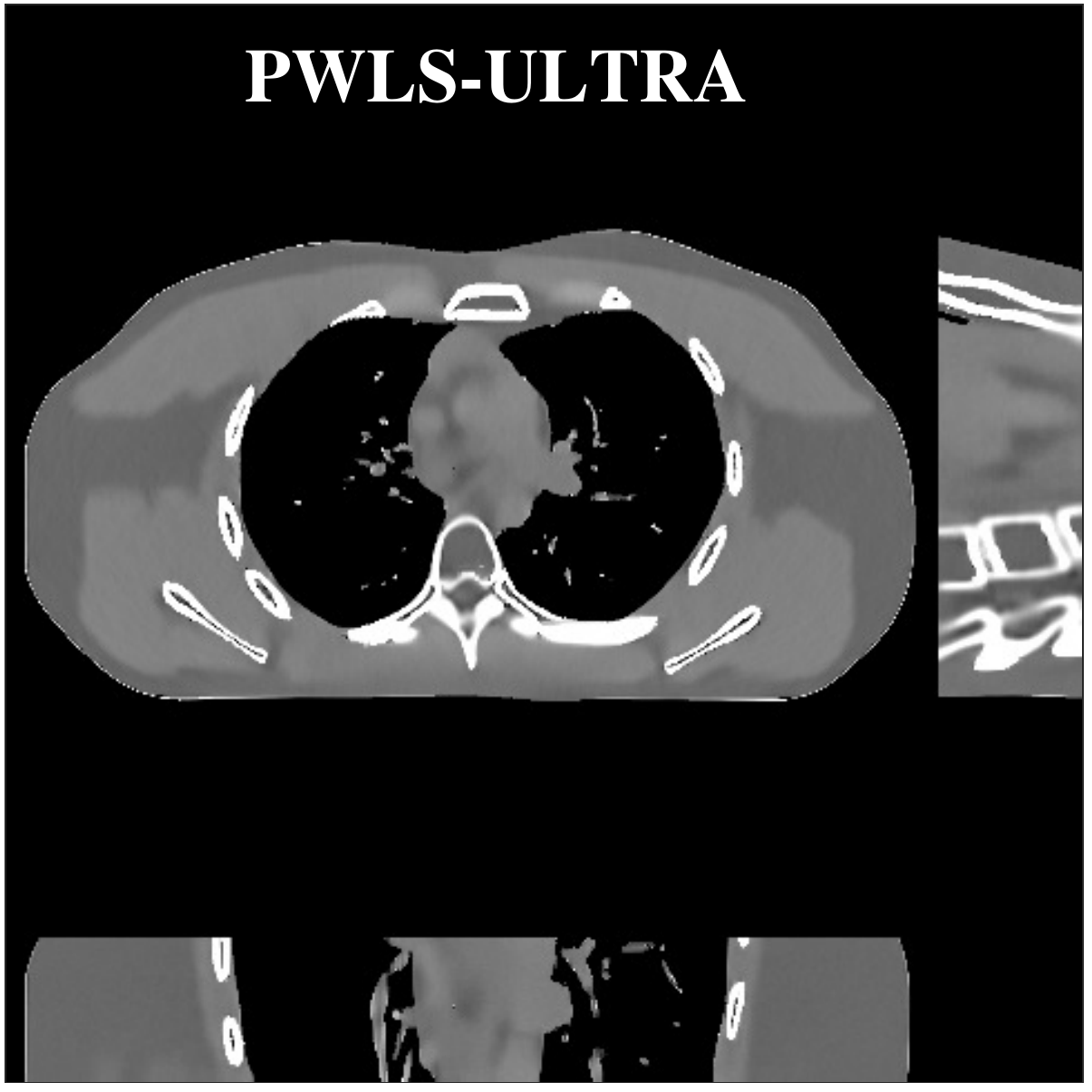}	};
	\spy on (0.87,0.12) in node [left] at (2.1,-1.2);		
	\spy on (-0.25,0.4) in node [left] at (2.1,1.65);		
	\end{tikzpicture}
	\caption{Cone-beam CT reconstructions (images from~\cite{zheng:18:pua}) of the XCAT phantom~\cite{segars:08:rcs} using the FDK, PWLS-EP~\cite{cho:15:rdf} (with edge-preserving regularizer), and PWLS-ULTRA~\cite{zheng:18:pua} ($K=15$) methods at dose $I_0 = 5 \times 10^3$ incident photons per ray, shown along with the ground truth (top left). The central axial, sagittal, and coronal planes of the 3D reconstruction are shown. The learning-based PWLS-ULTRA removes noise and preserve edges much better than the other schemes.}
	\label{fig:PWLSULTRA}
	\vspace{-0.15in}
\end{figure}

Recent works applied learned unions of transforms to other applications.
For example, the union of transforms model was pre-learned (from a dataset) and used in a clustering-based low-dose 3D CT reconstruction scheme~\cite{zheng:18:pua}.
Fig.~\ref{fig:PWLSULTRA} shows an example of high quality reconstructions obtained with this \add{scheme.}
While the work used a PWLS-type reconstruction cost, a more recent \add{method~\cite{yeravyongfes17}} 
\add{replaced} the weighted least squares data-fidelity term with 
\add{the}
shifted-Poisson likelihood penalty, which further improved image quality and reduced bias in the reconstruction in ultra low-dose \add{settings.}
\add{Other} 
recent \add{works} combined learned union of transforms models
with \add{material image models}
and applied it to image-domain material decomposition in dual-energy CT
with 
high quality \add{results~\cite{liravyongfes19,zlisaiyong19}.}

\subsubsection{Learning Structured Transform Models}

It is often useful to incorporate various structures and invariances in learning to better model natural data, and to prevent learning spurious features in the presence of noise and corruptions.
\add{Flipping and rotation invariant sparsifying transform learning} was recently
proposed and applied to image reconstruction in~\cite{wen2017frist}.
The regularization is similar to \eqref{unitemrireg}, but using $\W_k = \W \mathbf{\Phi}_k$ with a common \emph{parent transform} $\W$ and $\begin{Bmatrix}
\mathbf{\Phi}_k
\end{Bmatrix}$ denoting a set of known \addnew{flipping and rotation 
operators} that apply to each (row) atom of $\W$ and approximate \addnew{flips and rotations} 
by permutations (similar to \cite{qu2012undersampled}, but which used fixed 1D Haar wavelets as the parent).
This enables learning a much more structured but flexible
(depending on the number of \addnew{operators $\mathbf{\Phi}_k$}) model
than in \eqref{unitemrireg},
with clustering done more based on similar directional properties.
Images with more directional features are better modeled by 
\add{such learned transforms~\cite{wen2017frist}.}

\subsubsection{Learning Complementary Models -- Low-rank and Transform Sparsity}


\addnew{A recent work~\cite{wen2018power} proposed an approach called STROLLR (Sparsifying TRansfOrm Learning and  Low-Rank)} that combines two complementary regularizers:
one \addnew{exploiting (\emph{non-local}) self-similarity between regions,
and another exploiting transform learning} 
\add{that} is based on \emph{local} patch sparsity.
Non-local similarity and block matching models are well-known to have excellent performance in image processing tasks such as image denoising (with BM3D~\cite{dbov07}).
\add{The STROLLR} regularizer has the form $R(\x) = R_{1}(\x) + R_{2}(\x)$,
where the low-rank regularizer is as follows:
\begin{equation}
R_{1}(\x) = \min_{\begin{Bmatrix}
\mathbf{U}_j
\end{Bmatrix}} \sum_{j=1}^{N}\begin{Bmatrix}
\begin{Vmatrix}
\add{\mathcal{M}_{j} (\x)} - \mathbf{U}_{j}
\end{Vmatrix}_{F}^{2} + \eta^2 \, \text{rank}(\mathbf{U}_j)
\end{Bmatrix}, \label{strollrmrilrreg}
\end{equation}
and the transform learning regularizer is
\begin{align}
\nonumber R_{2}(\x) = & \min_{\W, \begin{Bmatrix}
\mathbf{z}_j
\end{Bmatrix}} \sum_{j=1}^{N}\begin{Bmatrix}
\begin{Vmatrix}
\W \mathbf{H}_{j} \x - \mathbf{z}_{j}
\end{Vmatrix}_{2}^{2} + \lambda^2 \, \begin{Vmatrix}
\mathbf{z}_{j}
\end{Vmatrix}_{0}
\end{Bmatrix} \\
& \;\;\;\;\;\; \text{s.t.} \;\; \W^{H}\W = \mathbf{I}. \label{strollrmrisreg}
\end{align}
Here, the \add{operator $\mathcal{M}_{j}$} is a block matching operator that extracts the $j$th patch $\P_{j} \x$ and the $L-1$ patches most similar to it and forms a matrix, whose columns are the $j$th patch and its matched siblings, ordered by degree of match. This matrix is approximated by a low-rank matrix $\mathbf{U}_j$ in \eqref{strollrmrilrreg}, with $\eta>0$.
The vector $\mathbf{H}_j \x$ is a vectorization of the submatrix that is the first $P$ columns \add{of $\mathcal{M}_{j} (\x)$}.
Thus the regularizer in \eqref{strollrmrisreg} learns a higher-dimensional 
\addnew{transform} (e.g., 3D transform for 2D patches), and jointly sparsifies non-local but similar patches.

\begin{figure}[!t] 
\begin{center} 
\begin{tabular}{cc}
\includegraphics[width=1.4in]{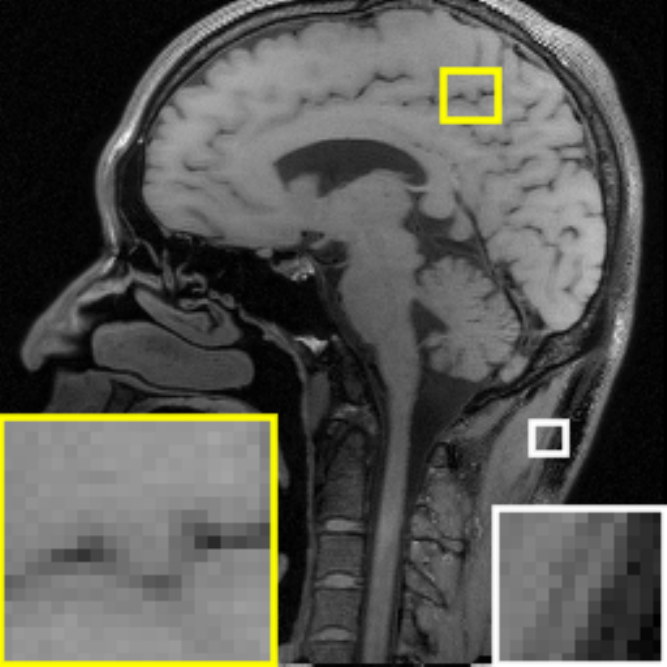} &
\includegraphics[width=1.4in]{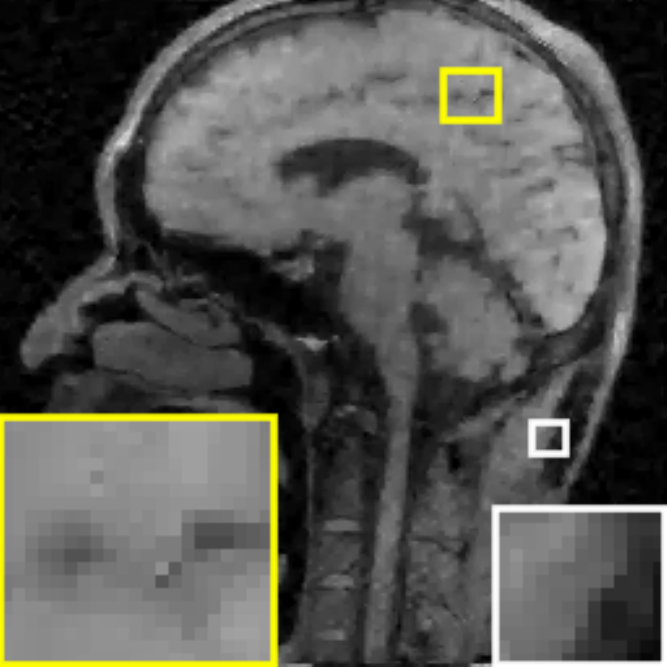} \\
{\footnotesize Ground Truth} & 
{\footnotesize Sparse MRI} \\
\includegraphics[width=1.4in]{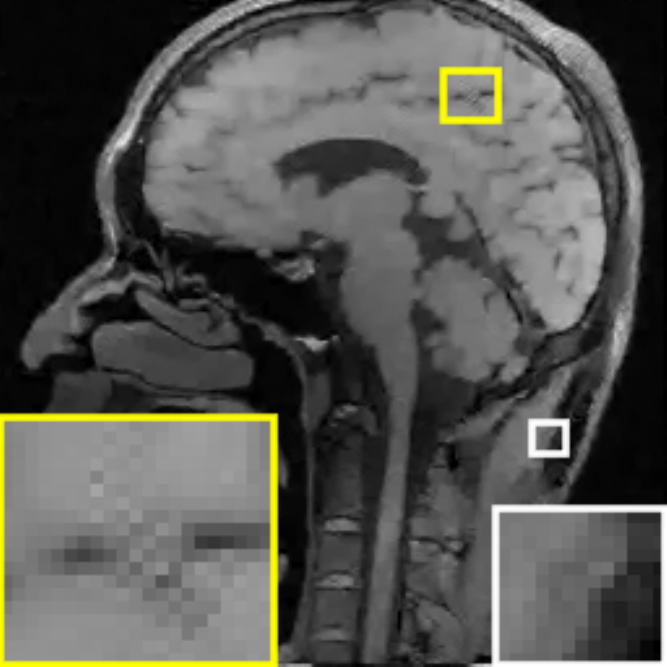} &
\includegraphics[width=1.4in]{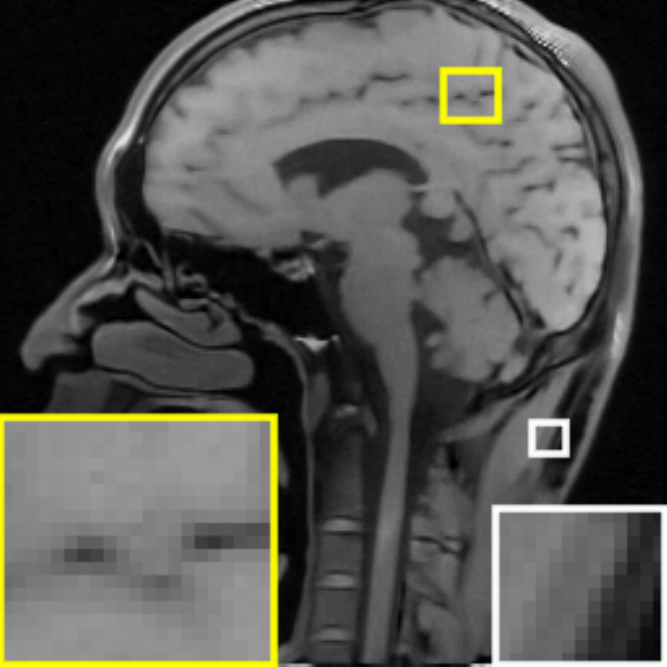} \\
{\footnotesize ADMM-Net} & 
{\footnotesize STROLLR-MRI } \\
\end{tabular} 
\caption{MRI reconstructions (images from~\cite{wensailukebres19}) with pseudo-radial sampling and 5x undersampling using Sparse MRI~\cite{lustig2007sparse} (PSNR = $27.92$ dB), ADMM-Net~\cite{sun2016deep} (PSNR = $30.67$ dB), and STROLLR-MRI~\cite{wen2018power} (PSNR = $31.98$ dB), along with the original image from~\cite{sun2016deep}. STROLLR-MRI clearly outperforms the nonadaptive Sparse MRI, while ADMM-Net also produces undesirable artifacts.} \label{fig:tlmri} 
\end{center} 
\vspace{-0.2in} 
\end{figure}

\add{STROLLR-MRI~\cite{wen2018power} was shown to achieve better CS MRI image quality over several methods including the supervised (deep) learning based ADMM-Net~\cite{sun2016deep}.}
Fig.~\ref{fig:tlmri} shows example MRI reconstructions and comparisons.
Similar to \add{UNITE-MRI,} 
there is an underlying grouping of patches,
but STROLLR-MRI exploits block matching and sparsity to implicitly \add{perform grouping.}


\subsection{Online Learning for Reconstruction} \label{seconlinelearning}

Recent works have proposed online learning of sophisticated models for reconstruction particularly of dynamic data from time-series 
\addnew{measurements~\cite{saibrianrajjeff17,briansairajjeff18,mardani16}.}
In this setting, the reconstructions are produced in a time-sequential manner from the incoming measurement sequence, with the models also adapted simultaneously and sequentially 
over time to track the underlying object's dynamics and aid reconstruction.
Such methods allow greater adaptivity to temporal dynamics
and can enable dynamic reconstruction with less latency, memory use, and computation
than conventional methods.
Potential applications include real-time medical imaging, interventional imaging, etc.,
or they could be used even for more efficient and \addnew{(spatially, temporally)} adaptive \addnew{\emph{offline} reconstruction
of large-scale (\emph{big}) data.}

\addnew{A recent work 
efficiently adapted} low-rank tensor models in an online manner for dynamic MRI~\cite{mardani16}.
Online learning for dynamic 
\add{image} reconstruction was shown to be promising in~\cite{saibrianrajjeff17,briansairajjeff18},
which adapted synthesis dictionaries to spatio-temporal patches.
\addnew{In this setup~\cite{briansairajjeff18},} 
measurements corresponding to a group
(called \emph{mini-batch}) of frames are processed at a time using a sliding window strategy.
The objective function for reconstruction is a weighted %
time average of instantaneous cost functions, each corresponding to a group of processed frames.
An exponential weighting (forgetting) factor for the instantaneous cost functions
controls the past memory in the objective.
The instantaneous cost functions include both a data-fidelity and a regularizer
(corresponding to patches in the group of frames) term.
The objective function thus changes over time and is optimized at each time point 
with respect to the most recent mini-batch of frames and corresponding sparse coefficients (with older frames and coefficients fixed),
but the dictionary is itself adapted therein to all the data. 
Each frame can be reconstructed from multiple overlapping temporal windows
and a weighted average of those used as the final estimate.

The online learning algorithms in~\cite{briansairajjeff18} achieved 
\add{computational efficiency}
by using warm start initializations (that improve over time)
for variables and frames based on estimates in previous windows,
and thus running only a few iterations of optimization for each new window.
They stored past information in small (cumulatively updated) matrices for the dictionary update \add{(low memory usage)}. 
The \addnew{methods were} 
significantly more efficient and more effective
than batch learning-based techniques for dynamic MRI
that iteratively learn and reconstruct from \emph{all} k-t space measurements.
Given the potential of online learning methods to transform dynamic and large-scale imaging,
we expect to see growing interest and research in this domain.

\subsection{Connections between Transform Learning Approaches and Convolutional Network Models}
\label{secfilterbanks}

The sparsifying transform models in Section~\ref{sec:datadriven:transformlearning}
have close connections with convolutional filterbanks.
This subsection and the next review some of these connections and implications for reconstruction. 

\subsubsection{Connections to Filterbanks}

Transform learning and its application
to regularly spaced image
\addnew{patches~\cite{pfisbres19}}
can be equivalently performed using convolutional operations.
For example, applying an atom of the transform
to all the overlapping patches of an (2D) image via inner products
is equivalent to convolving the image
with a \emph{transform filter} that is the (2D) flipped version of the atom.
Thus, sparse coding in the transform model
can be viewed as convolving the image with a set of transform filters
(obtained from the transform atoms)
and thresholding the resulting filter coefficient maps,
and transform learning can be viewed as equivalently 
learning convolutional sparsifying filters
\cite{chun:18:cao-asilomar,chun:18:cao-arxiv}.
When using only a regularly spaced subset of patches,
the above interpretation of transform sparse coding modifies to convolving the image with the transform filters,
downsampling the results, and then thresholding~\cite{pfisbres19}.
Transform models 
based on clustering~\cite{sravTCI1}
add 
non-trivial complexities to this process. 

Applying the matrix $\W^{H}$ 
to the sparse codes of all overlapping patches
and spatially aggregating the results,
an operation used in iterative transform-based reconstruction algorithms~\cite{sravTCI1},
is equivalent to filtering the thresholded filter coefficient maps
with corresponding matched filters
(complex conjugate of transform atoms)
and summing the results over the channels.
These equivalences between patch-based and convolutional operations
for the transform model
contrast with the case for the synthesis dictionary model
in Section~\ref{sec:datadriven:dictionarylearning},
where the patch-based and convolutional versions of the model are not equivalent in general.
When a disparate set of (e.g., randomly chosen) image patches or operations
such as block matching~\cite{wen2018power}, etc., are used with the transform model,
the underlying operations do not correspond to convolutions
(thus, the 
transform learning frameworks can be viewed as more general).
Typically the convolutional implementation of transforms
is more computationally efficient than the patch-based version
for \addnew{large filter sizes~\cite{pfisbres19}.}


Recent works have exploited the filterbank interpretation
of the transform model~\cite{pfisbres19,ravacfess18,saibrenmulti}.
\addnew{For example, \cite{pfisbres15}
learned filterbanks for MRI reconstruction.
In 
~\cite{pfisbres19},
the authors studied} alternative properties and regularizers for transform learning.

\subsubsection{Multi-layer Transform Learning}

\addnew{A recent work~\cite{saibrenmulti} proposed}
learning multi-layer extensions of the transform model
(dubbed deep residual transforms (DeepResT))
\add{that} 
mimic convolutional neural networks (CNNs)
by incorporating components such as
filtering, nonlinearities, pooling, 
and stacking;
however, the learning was done using unsupervised
\emph{model-based transform learning-type cost functions}. 

\begin{figure*}[!t]
\hspace{-0.0in}\includegraphics[width=1.0\textwidth]{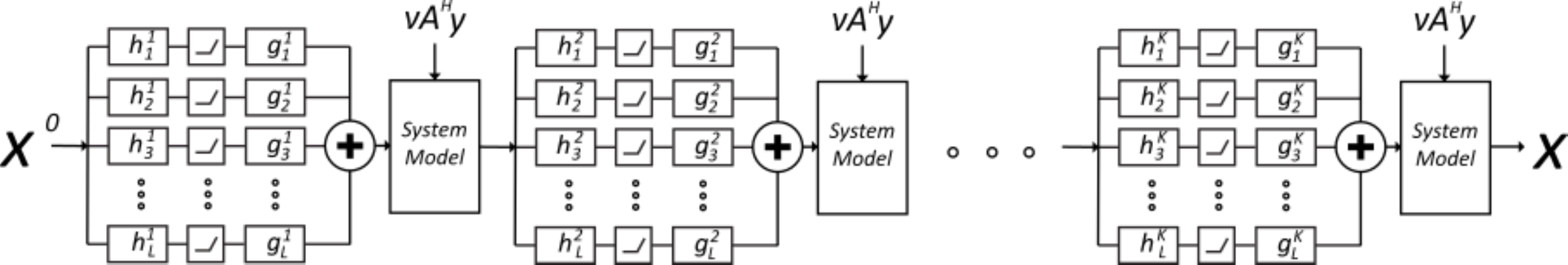}
\caption{The reconstruction model (see~\cite{ravacfess18}) derived from the image update step of \add{the square transform learning-based image reconstruction algorithm in~\cite{sravTCI1}.} 
The model here has $K$ layers corresponding to $K$ iterations. Each layer first has a decorruption step that computes the second term in \eqref{imupd} using filtering and thresholding operations, assuming a transform model with $L$ filters. This is followed by a system model block that adds the fixed bias term $\nu \A^{H} \y$ to the output of the decorruption step and performs a least-squares type image update (e.g., using CG) to enforce the imaging forward model.}
\label{fig:physicsdrivenlearning}
\vspace{-0.04in}
\end{figure*}


In the conventional transform model,
the image is passed through a set of transform filters and thresholded (the non-linearity)
to generate the sparse coefficient maps.
\add{In} the DeepResT model,
the residual (difference) between the filter \add{outputs} 
and their \add{sparse} 
versions is computed
and \add{these} 
residual maps for different filters 
are stacked together to form a residual volume
\add{that is} jointly sparsified in the next layer. 
\add{To} prevent dimensionality explosion,
each filtering of the residual volume in the second and subsequent layers
produces a 2D output (for a 2D initial image).
The multi-layer model \add{thus} consists of successive joint sparsification of residual maps several \add{times (cf. Fig. 1 in \cite{saibrenmulti} and Fig. 9 in \cite{wensailukebres19}).}
\add{The filters and sparse maps} 
\add{in all layers of the (\emph{encoder}) network are
jointly and efficiently learned in~\cite{saibrenmulti} from images 
to provide the smallest sparsification 
residuals in the final (output) layer,
a transform learning-type cost.}
The learned model and multi-layer sparse coefficient maps can then be backpropagated
in a linear fashion (\emph{decoder})
to generate 
\add{image approximations.}
The DeepResT model also downsampled (pooled) the residual maps \add{(along the filter channel dimension)} in each encoder layer
before further filtering them,
providing robustness to noise and data corruptions.
\add{The learned models were shown~\cite{saibrenmulti} to provide promising performance
for denoising images when learning directly from noisy data,}
and moreover learning stacked multi-layer encoder-decoder modules was shown to improve performance,
especially at high noise levels.
Application of such deep transform models to medical image reconstruction
is an \add{ongoing area~\cite{zhengsaietalmulti2019}} of potent research.

\subsection{Physics-Driven Deep Training of Transform-Based Reconstruction Models} \label{secphysicsdrivendeep}

There has been growing recent interest in supervised learning approaches
for image reconstruction~\cite{schlemper18}.
These methods learn the parameters of reconstruction algorithms from training datasets
(typically consisting of pairs of ground truth images and initial reconstructions from measurements)
to minimize the error in reconstructing the training images
from their typically limited or corrupted measurements.
For example, the reconstruction model can be a deep CNN
(typically consisting of encoder and decoder parts)
that can be trained (as a denoiser)
to produce a reconstruction from an initial corrupted version~\cite{leeye18}.
Section~\ref{sec:deepmodel} discusses such approaches in more detail.
These methods can often require large training sets
to learn billions of parameters (e.g., filters, etc.).
Moreover, learned CNNs (\emph{deep learning})
may 
not typically or rigorously incorporate the imaging measurement model
or the information about the Physics of the imaging process,
which are a key part 
of solving inverse problems.
Hence, there has been recent interest
in learning the parameters of iterative algorithms
that solve regularized inverse problems~\cite{sun2016deep,ravchfess17}
(cf. Section~\ref{sec:deepmodel} for more such methods).
These methods can also typically have fewer free parameters to train.

\add{Recent works have interpreted early transform-based BCS algorithms
as deep physics-driven convolutional networks learned on-the-fly,
i.e., in a blind manner, from measurements~\cite{ravchfess17,ravacfess18}.}
For example, the image update step in the \add{square} transform BCS \add{(that learns a unitary transform)} algorithm \add{in~\cite{sravTCI1}} 
involves a least squares-type optimization with the following normal equation:
\begin{align}
\G \x^{k}  =  \nu \A^{H}\y + \sum_{j=1}^{N} \P_{j}^{T}\D^{k}\H_{\lambda}(\W^{k}\P_j \x^{k-1}), \label{imupd}
\end{align}
where $\nu=1/\beta$
(for $\beta$ in Section~\ref{sec:datadriven:dictionarylearning}
or \ref{sec:datadriven:transformlearning})
and $k$ denotes the iteration number in the block coordinate descent \add{reconstruction algorithm}. 
Matrix $\D^{k}\triangleq\begin{pmatrix}
\W^{k}
\end{pmatrix}^H$ is a (matched) synthesis operator,
and $\G \triangleq \sum_{j=1}^{N} \P_{j}^{T}\P_{j} +  \nu \A^{H}\A$
is a fixed matrix.
The hard-thresholding in~\eqref{imupd}
corresponds to the solution of the sparse coding step \add{of the BCD algorithm~\cite{sravTCI1}.} 

Fig.~\ref{fig:physicsdrivenlearning} shows an \emph{unrolling} of $K$ iterations (\emph{layers})
of \eqref{imupd},
with fresh filters in each iteration.
Each layer has a system model block that solves \eqref{imupd} (e.g., with FFTs or CG),
whose inputs are the two terms on the right hand side of \eqref{imupd}:
the first term is a \emph{fixed bias} term;
and the second term (denotes a decorruption step) is computed via convolutions by first applying the transform filters
(denoted by $h_{l}^{k}$, $1\leq l \leq L$ in Fig.~\ref{fig:physicsdrivenlearning})
followed by thresholding (the non-linearity)
and then matched synthesis filters (denoted by $g_{l}^{k}$, $1\leq l \leq L$),
and summing the outputs over the filters. %
This is clear from writing the second term in~\eqref{imupd} as $\sum_{l=1}^{L}\sum_{j=1}^{N} \P_{j}^{T}\mathbf{d}_{l}^{k}\H_{\lambda}(\mathbf{r}_{l}^{k^T}\P_j\x^{k-1})$, with $\mathbf{d}_{l}$ and $\mathbf{r}_{l}$ denoting the $l$th columns of $\D$ and $\mathbf{R}=\W^{T}$, respectively.
Each of the $L$ terms forming the outer summation here
corresponds to the output of an arm (of transform filtering, thresholding, and synthesis filtering) in the decorruption module of Fig.~\ref{fig:physicsdrivenlearning}.
\add{Since the BCS scheme in~\cite{sravTCI1} does not use training data, but rather learns the transform filters as part of the iterative BCD algorithm (hence, the transform could change from iteration to iteration or layer to layer), it can be interpreted as learning the model in Fig.~\ref{fig:physicsdrivenlearning} in an
on-the-fly sense from measurements.}



Recent works~\cite{ravchfess17,ravacfess18} learned the filters in this multi-layer model
(a block coordinate descent or BCD Net \cite{chfess18})
with soft-thresholding ($\ell_1$ norm-based) nonlinearities and trainable thresholds
using a greedy scheme to minimize the error in reconstructing a training set
from limited measurements.
These and similar approaches
(including the transform-based ADMM-Net~\cite{sun2016deep})
involving unrolling of \add{typical
image reconstruction algorithms}
are 
\emph{physics-driven deep training} methods
due to the systematic inclusion of the imaging forward model in the convolutional network.
Once learned, the reconstruction model can be efficiently applied to test data
using convolutions, thresholding, and least squares-type updates.
While \cite{ravchfess17,ravacfess18} did not
enforce the corresponding synthesis and transform filters
(in each arm of the decorruption module) to be matched in each layer,
recent work \cite{chfess18} learned matched filters,
improving image quality.
The learning of such \add{physics-driven} 
\add{networks}
is an active area of research,
with interesting possibilities for new innovation
in the convolutional models in the architecture
motivated by more recent transform and dictionary learning based \addnew{(or other)} reconstruction methods.
In \addnew{such} methods,
the thresholding operation
is the key to exploiting sparsity.

\section{Deep Learning Methods}
\label{sec:deepmodel}

One of the most important recent developments
in the field of image reconstruction
is the introduction of deep learning approaches
\cite{wang:18:iri}. 
Motivated by the tremendous success of deep learning
for image classification\cite{krizhevsky2012imagenet,he2016deep},
image segmentation\cite{ronneberger2015u}, 
denoising~\cite{zhang2017beyond}, etc, 
many groups 
have recently
successfully applied
deep learning approaches
to various image reconstruction problems
such as \add{in}
X-ray CT \cite{kang2017deep,chen2017low,kang2018deep,adler2018learned,wolterink2017generative,jin2017deep,han2017framing},
MRI \cite{wang2016accelerating,hammernik2018learning,lee2018deep,han2017deep,jin2017deep,schlemper18,zhu2018image},
PET \cite{gong2018iterative,gong2018pet}
ultrasound \cite{luchies2018deep,yoon2019efficient},
and optics~\cite{rivenson2017deep,nehme2018deep,sinha2017lensless}.
\addnew{
As of mid 2019,
two commercial CT vendors
received FDA approval
for deep learning image reconstruction
\cite{fda:19:canon-aice}
\cite{fda:19:ge-dlir}.
}

The sharp increase
in deep learning approaches \add{for} image reconstruction \add{problems} 
\add{may} be due to the ``perfect storm'' 
resulting from a combination of multiple attributes in perfect timing:
availability of 
large public data,
well-established GPU infrastructure in the image reconstruction community,
easy-to-access deep learning toolboxes, industrial push, and open publications using arXiv, etc.

For example, one 
important public data set
that has significantly contributed to this wave
is the 2016 American Association of Physicists in Medicine (AAPM)
Low-Dose X-ray CT Grand Challenge data set~\cite{mccollough2017low}.
The training data sets consist of normal-dose and quarter-dose abdominal CT data from ten patients.
Another emerging important data set
is the fast MRI data set by NYU Langone Health and Facebook~\cite{zbontar2018fastmri}.
The dataset comprises raw k-space data
from more than 1,500 fully sampled knee MRIs
and DICOM images from 10,000 clinical knee MRIs
obtained at 3 Tesla or 1.5 Tesla.  

In addition, 
GPU methods have
been extensively implemented to 
accelerate iterative methods in the field of image reconstruction.
As a result,
open deep learning toolboxes such as Tensorflow, pyTorch, MatConvNet, etc.,
based on GPU programming,
are easily accessible to researchers in the field of image reconstruction.
\add{Moreover,} 
\add{the} industry has been engaging in 
a big push in this development from the early phase,
since deep learning based image reconstruction methods are well-suited to their business models.
This is because  the training can be done by the vendors with large databases
and the users could
enjoy high quality reconstruction results at near real-time reconstruction speed.

Given the relatively long publication cycle
for regular journals in the field of image reconstruction,
most new developments are found in arXiv preprints,
well before they are formally accepted by the journals.
This new trend of open publication
facilitates the significant progress in this area within a short time period.
The following subsection
reviews the recent developments
based on the peer-reviewed publications and arXiv preprints.

\subsection{Categories of the existing approaches}

\begin{figure}[ht!]
	\center
	\subfigure[Image-domain Learning]{\includegraphics[width =0.4\textwidth]{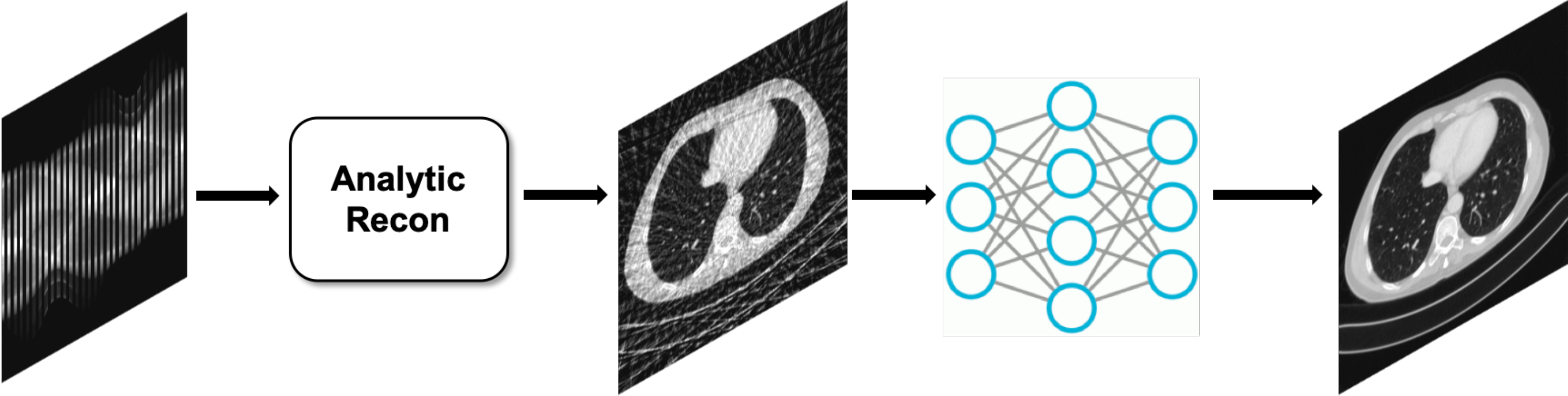}}~~\\
	\subfigure[Hybrid-domain Learning]{\includegraphics[width =0.4\textwidth]{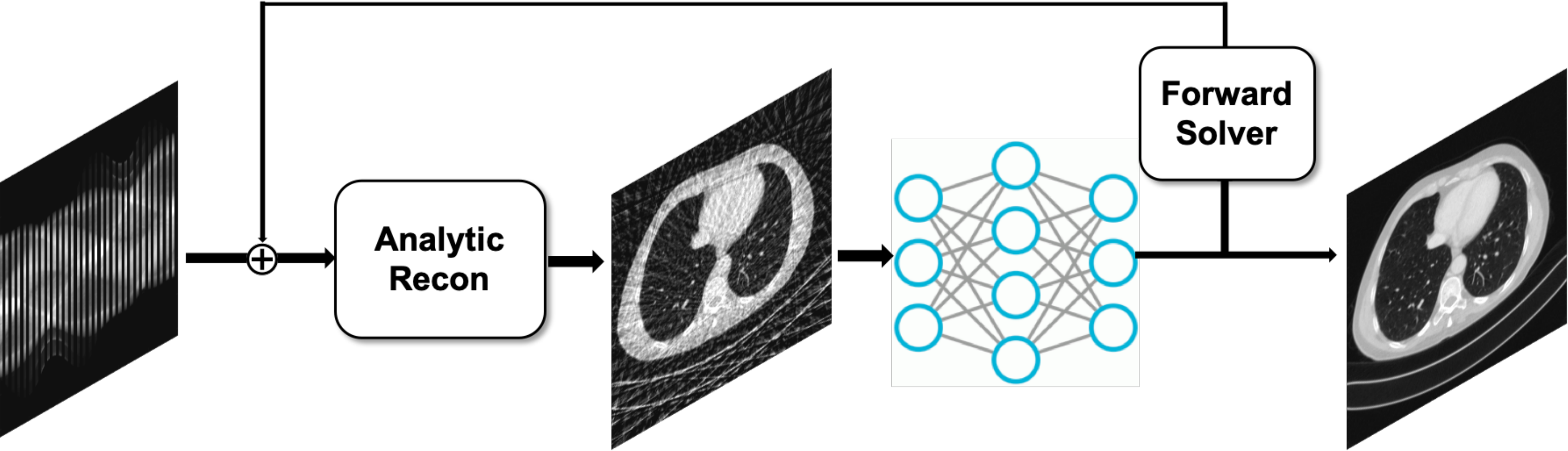}}~~\\
	\subfigure[AUTOMAP]{\includegraphics[width =0.4\textwidth]{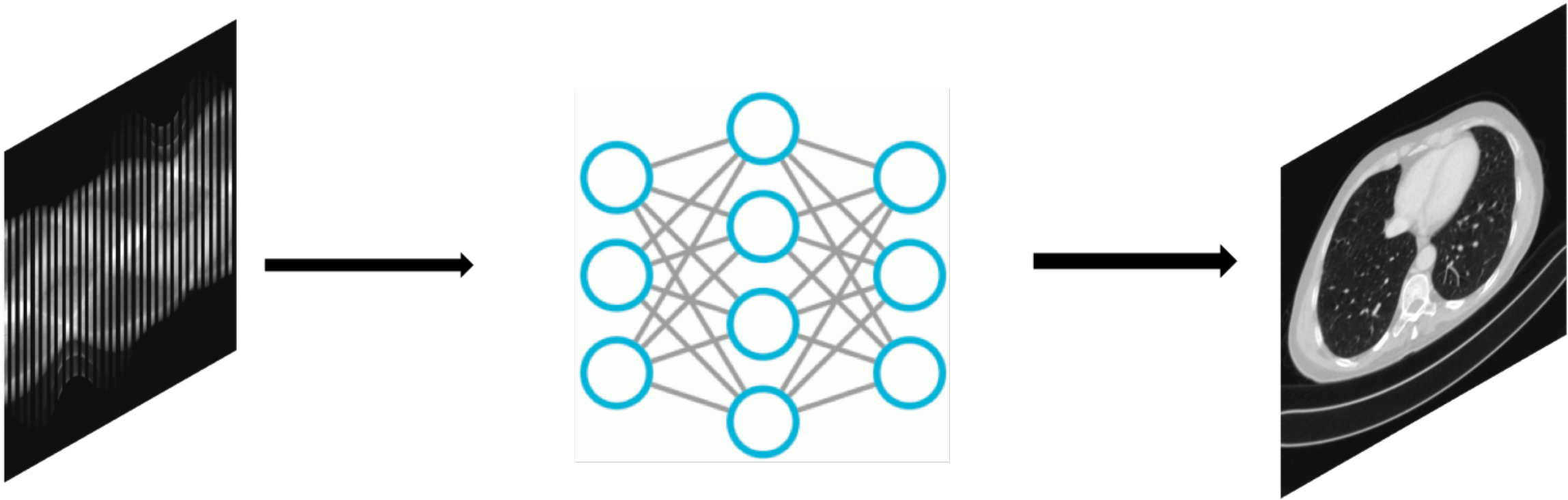}}~~\\
	\subfigure[Sensor-domain Learning]{\includegraphics[width =0.4\textwidth]{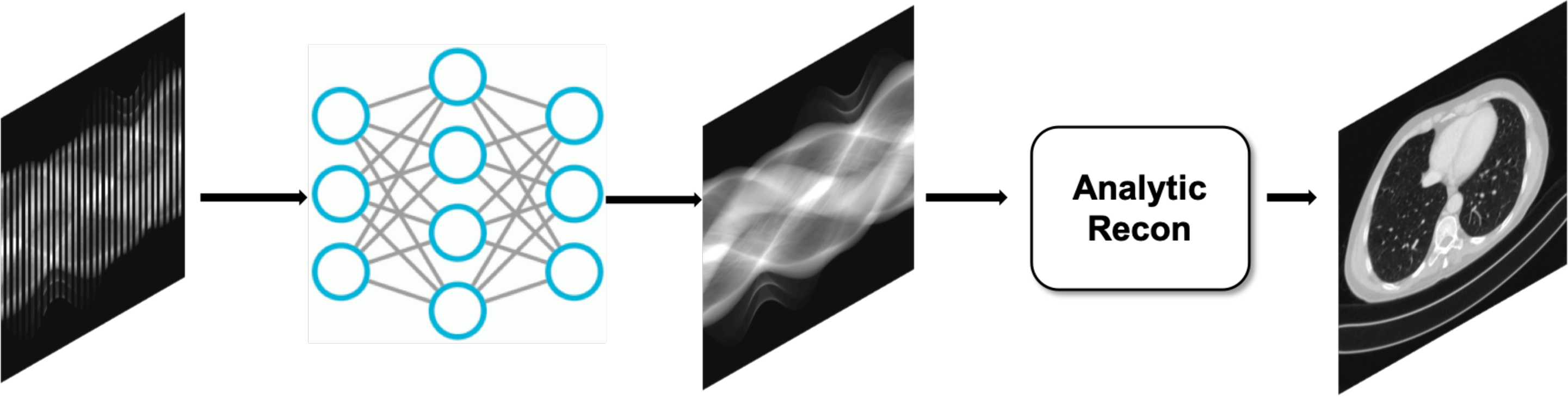}}~
	\caption{Various realizations of deep learning for image reconstruction. }
	\label{fig:architecture}
\end{figure}

This section starts with reviews  of various
network architectures \add{and design principles} for image reconstruction problems.
Fig.~\ref{fig:architecture}
illustrates typical architectures at a high level, which are commonly used in the literature.
\add{However, this field is rapidly growing, so a complete review of the design principles is beyond the scope of the paper.}


\subsubsection{Image-domain learning}

In image domain approaches
\cite{hammernik2018learning,kwon2017parallel,kang2017deep,chen2017lowBOE,jin2017deep,han2017deep,han2017framing,chen2017low,wolterink2017generative},
artifact-corrupted images are first generated
from the measurement data using some analytic methods
(e.g., FBP, Fourier transform, etc.),
from which neural networks are trained to learn the artifacts
(see Fig.~\ref{fig:architecture}(a)).  
For example, the low-dose and sparse CT neural networks
\cite{kang2017deep,chen2017lowBOE,jin2017deep,han2017framing,chen2017low,wolterink2017generative}
belong to this class,
where the noise corrupted images are first generated
from the noisy or sparse view sinogram data using FBP,
after which the artifacts are learned by comparing with the noiseless label images.
In MR applications,
the early U-Net architectures for compressed sensing MRI
\cite{lee2017deep,han2017deep}
were also designed to remove the aliasing artifacts
after obtaining the Fourier inversion image from the downsampled k-space data.


\add{
In particular, FBPConvNet~\cite{jin2017deep} showed that  image domain
networks can be derived by unrolling sparse recovery 
for one specific class of inverse problems:
those where the normal operator associated with the forward model
is a convolution.
For this class of inverse problems,
a CNN
then emerges.
This class of normal operators 
includes MRI,
parallel-beam
X-ray CT, and diffraction tomography (DT).}

A current trend in image domain learning
is to use more sophisticated loss functions to overcome the observed
smoothing artifacts.
For example, in~\cite{yang2018low},
the authors used the perceptual loss and Wasserstein distance loss
to improve the resolution.

\subsubsection{Hybrid-domain learning}

In this class of approaches
\cite{hammernik2018learning,sun2016deep,wang2016accelerating,sun2016deep,schlemper18,gupta2017cnn,adler2018learned,lee2018deep,kang2018deep,quan2018compressed,chen2018learn,aggarwal2017modl,wu2017iterative,gong2017iterative},
the data consistency term is imposed in the neural network training and inference
to improve the performance as shown in Fig.~\ref{fig:architecture}(b).
The physics-driven deep training methods in Section~\ref{secphysicsdrivendeep}
included the full data-fidelity based image update in each layer.


\add{
The Learned ISTA (LISTA) \cite{gregor2010learning} is one of the earliest unfolding approaches that
 uses a time unfolded version of the iterative soft-threshodling (ISTA) \cite{beck2009fast}. 
 Specifically, the weight matrices  and  sparsifying soft-thresholding operator
are learned from the  data (as also done in later works discussed in Section~\ref{secphysicsdrivendeep}).
%
 }

\add{The variational neural network for compressed sensing MRI
\cite{hammernik2018learning}
derived an unrolled neural network
that uses a data consistency term for each layer.
%
Specifically,
the variational network is based on unfolding
the following optimization problem:
\begin{eqnarray}
\min_{\x} \frac{\lambda}{2}\|\y-\A\x\|^2 +  \Rc(\x),
\end{eqnarray}
where the regularization term $\Rc(\x)$ is represented as a sum of multichannel operations:
\begin{eqnarray}
\Rc(\x) = \sum_{i=1}^M \langle \Phi_i(\blmath{K}_i \x), \mathbf{1} \rangle,
\end{eqnarray}
where $\blmath{K}_i$ denotes the $i$th linear operator
represented by the $i$th channel convolution
(like the transform filters
in Section~\ref{sec:datadriven}),
and $\Phi_i$ denotes the associated activation function.
In a variational network~\cite{hammernik2018learning},
the convolution-based linear operator $\blmath{K}_i$,
the gradient of the activation function $\Phi_i'$,
and the regularization parameter $\lambda^{k}$
are learned for each unfolded \addnew{Landweber iteration steps}.
Related 
approaches have been taken in dynamic cardiac MRI
\cite{schlemper18}.}

In ADMM-Net~\cite{sun2016deep},
the unrolled steps of the  alternating direction method of multipliers (ADMM) 
based reconstruction
algorithm
are mapped to each layer of a deep neural network.
Recently,
the primal-dual algorithm was extended
to obtain a CNN-based learned primal-dual approach
\cite{adler2018learned}. 
In this approach,
two neural networks are learned for the primal step and dual step. 
The projected gradient method
has also been extended
to a neural network approach
\cite{gupta2017cnn}.

\add{Another class of popular hybrid domain approaches
is based on the CNN penalty and plug-and-play model.
Specifically, 
in CNN penalty approaches~\cite{wang2016accelerating},
a neural network is used as a prior model
within an 
MBIR framework.
Rather than using a CNN penalty explicitly, 
in the plug-and-play approach~\cite{gong2017iterative,wu2017iterative}, 
the denoising step of an iteration like ADMM 
is replaced with a neural network denoiser.
Similarly, 
the deep image prior approach~\cite{ulyanov2018deep}
formulates the image reconstruction problem as
\begin{eqnarray}
\min_{\thetab} & \frac{\lambda}{2}\|\y-\A\z(\thetab)\|^2 +  \Rc(\z(\thetab))\\
\mbox{subject to} & \z(\thetab) = \blmath{G}_{\thetab}(\mathbf{v}),
\end{eqnarray}
where $\blmath{G}_\thetab$ is a deep neural network parameterized by $\thetab$. \add{The dimension
of the parameter $\thetab$ is usually determined by the neural network architecture.}
In the original deep prior model~\cite{ulyanov2018deep},
the input ($\mathbf{v}$) for the neural network was a noise vector.
Instead, in its recent application to PET image reconstruction~\cite{gong2018pet}, 
simultaneously acquired MRI data  was used
as the input to the neural network for PET reconstruction using deep image prior.}


\subsubsection{AUTOMAP}

The Automated Transform by Manifold Approximation
(AUTOMAP)~\cite{zhu2018image}
(see Fig.~\ref{fig:architecture}(c)) approach
learns a direct mapping
from the measurement domain to image domain using a neural network.
This approach requires a fully connected layer followed by convolution layers,
leading to high
memory requirements
for storing that fully connected layer,
currently limiting
AUTOMAP applications to small size reconstruction problems \add{in MRI}.

\subsubsection{Sensor-domain learning}

Sensor-domain learning approaches
try to learn the sensor domain interpolation and denoising
using a neural network as shown in Fig.~\ref{fig:architecture}(d).
For low-dose CT,
\cite{wurfl2016deep} designed a neural network in the projection domain,
yet the neural network is trained 
in an end-to-end manner
from the sinogram to the image domain.
Accordingly, the final output of the neural network
is a data-driven ramp filter
designed by minimizing the image domain loss.
\add{
Metal artifact correction in CT
is another opportunity for sensor-domain learning
\cite{ghani:19:fac,ghani2018deep}. 
For low-dose CT and sparse view CT, direct
sinogram domain processing using deep neural network was also proposed \cite{lee2018deep_sinogram,ghani2018deep_low}.
}%
In k-space deep learning for accelerated \add{MRI~\cite{han2018k,lee2018k,akcakaya19}},
neural networks were designed to learn
k-space interpolation kernels in an end-to-end manner
from k-space to the image domain using an image domain loss.

\subsubsection{Some variations}

\begin{figure}[!t]
	\center
	{\includegraphics[width=\linewidth]{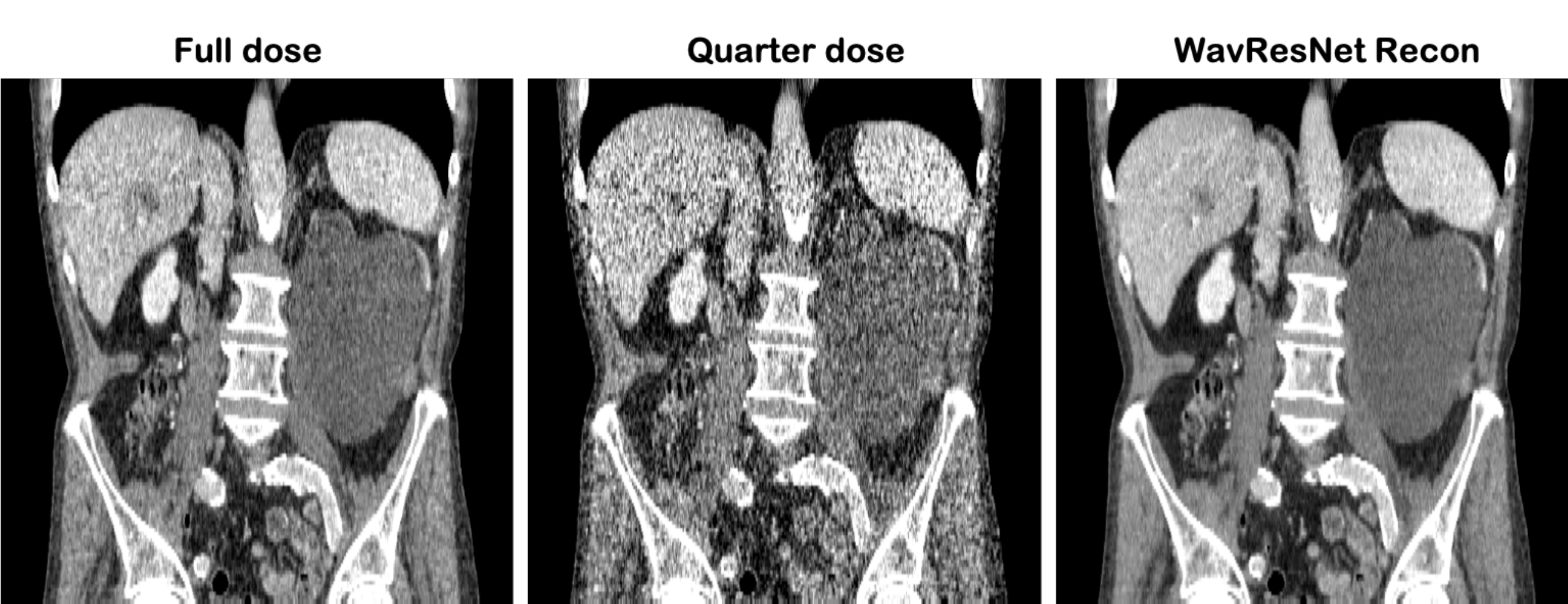}}
	\caption{
	\add{
	Left to right: full-dose FBP reconstruction,
	quarter-dose FBP reconstruction,
	and
	WavResNet denoising results \cite{kang2018deep}
	applied to 25\% dose FBP images.
	The detailed textures, vessel structure,
	and cancer lesions are clearly seen in the denoised results.}
	}
	\label{fig:wavresnet}
	\vspace{-0.15in}
\end{figure}

In \cite{kang2017deep},
the neural networks were designed to learn the relationship
between contourlet transform coefficients
of the low-dose input and high dose label data.
Later, this problem is formally extended to wavelet domain residual network (WavResNet) 
to improve the performance \cite{kang2018deep} (see Fig.~\ref{fig:wavresnet} for an example).
\addnew{Here, the choice of appropriate  transform domain facilitating efficient learning is important and usually based on domain expertise.}
For example, in a recent deep neural network architecture
for interior tomography problems,
\cite{han2018one}
observed that the neural network is more robust
with respect to different ROI sizes, detector pitch, short scan and sparse view artifacts,
if the neural network is designed
in the differentiated backprojection (DBP) domain.
The DBP is well-known in \add{the} CT community
for its robustness to short scan artifact,
interior tomography, etc, \addnew{which clearly shows the importance of domain expertise in designing neural networks.}

\subsection{Semi-supervised and Unsupervised Learning}

Most deep learning approaches for image reconstruction
have been based on the supervised learning framework.
For example,
in the low-dose CT reconstruction problems,
the neural network is trained
to learn the mapping between the noisy image and the noiseless (or high dose) label images.
Similar approaches are taken in accelerated MRI,
where the relationship between \add{highly} accelerated and artifact corrupted input
and the fully sampled label data are learned using training data.

Unfortunately,
in many imaging scenarios
the noiseless label images are difficult to obtain or even impossible to acquire.
For example, in low-dose CT problems,
an institutional reviewer board (IRB) rarely approves experiments
that would require two exposures at low and high dose levels
due to the potential risks to patients.
This is why in the AAPM X-ray CT Low-Dose Grand Challenge,
the matched low-dose images were generated
by adding synthetic noise to the full dose sinogram data.
Even in accelerated MRI,
high-resolution fully sampled k-space data is very difficult
to acquire
due to the long scan time,
and impossible to collect
for dynamic MRI data sets
that are all inherently under-sampled.
Therefore,
neural network training without reference
or with small reference pairs
are very important in the field of image reconstruction.

One of the earliest works in this regard
was a low-dose CT denoising network
\cite{wolterink2017generative}.
Instead of using matched high-dose data,
the authors employ the GAN loss to match the probability distribution.
One of the limitations of this work
is that the network is very sensitive and,
without careful training, spurious artifacts are often generated
due to the generative nature of GAN.
To address this problem,
a cycleGAN architecture
employed cyclic loss and identity loss
for multiphase cardiac CT problems
\cite{kang2019cycle}.
Thanks to \addnew{the two loss functions},
the authors demonstrated that no spurious artifact appeared in their results
even without reference data.

Besides, several of the learning approaches in Section~\ref{sec:datadriven} also do not require reference data \add{(or can use limited reference data)} to provide high quality reconstructions.


\subsection{Interpretation of Deep Models}

One of the major hurdles of the deep learning approaches for image reconstruction
is the black-box nature of neural networks.
This is especially problematic for medical imaging applications,
since many clinicians are concerned about whether
the performance improvement is real or cosmetic. 

\addnew{
The modern reconstruction techniques,
such as compressed sensing,
can be considered as {\em representation learning} methods
that aim at finding the optimal parsimonious representation
under the data fidelity term.}
\add{
Unfortunately, the classical  approaches
usually require computationally expensive optimization methods
to find the optimal representation. 
One of the important take-home messages of this section
is to show that deep learning approaches are indeed
another form of representation learning approaches,
which have advantages compared to the classical approaches.
In the following, we start to revisit the classical approaches in this aspect.
 }
 
\subsubsection{\add{Image Reconstruction via Representation Learning}}

\add{
One can formulate image reconstruction problems as
\begin{eqnarray}
\min_{\x \in \Xbc} &\|\y-\A\x\|^2, 
\end{eqnarray}
where $\Xbc$ denotes the low dimensional manifold where the unknown $\x$ lives.
For example, 
 $\Xbc$ can be represented by
 \begin{eqnarray}\label{eq:Xc}
 \Xbc = \left\{ \x ~|~ \x=\sum_{i\in I} \langle {\blmath b}_i, \x \rangle \tilde  {\blmath b}_i,\right\}
 \end{eqnarray}
 where $I\subseteq \mathbb{N}$ is an index set
 (for example, in compressed sensing, $I$ is a sparse index set such that the signal can be represented as a sparse combination of basis elements.)
Here, a family of functions $\{{\blmath b}_i\}_{i\in \mathbb{N}}$ is usually selected as
a {\em frame} that satisfies the following equality~\cite{duffin1952class}:
 \begin{eqnarray}\label{eq:framebound}
A\|\x\|^2 \leq   \sum_{i\in \mathbb{N}} |\langle{\blmath b}_i, \x \rangle|^2 \leq B\|\x\|^2, \quad \forall \x \in H, 
\end{eqnarray}
where $A,B>0$ are called the frame bounds, and $H$ denotes the specific Hilbert space. If $A=B$, then the frame is called a tight frame.
In \eqref{eq:Xc}, another family of functions $\{\tilde{\blmath b}_i\}_{i\in \mathbb{N}}$ form the {\em dual} frame
satisfying the equality:
$$\sum_{i\in \mathbb{N}}  \langle{\blmath b}_i, \tilde{\blmath b}_i \rangle = 1. $$
If the frame basis is chosen in a multiresolution manner, it is called {\em framelet}.
In compressed sensing, 
the frames are usually chosen from wavelet transforms, learned dictionaries, or other redundant bases,
where the sparse combination of a subset of the frame can represent the unknown signals with high accuracy.
 }

\add{Even for recent approaches such as low-rank Hankel structured matrix completion approaches, the same frame interpretation exists.
Specifically, for a given low-rank Hankel matrix $\hank_{[d]}^{[n]}(\x)$,
let $\Phib =[\phib_1,\cdots,\phib_n] \in \Rd^{n\times n}$ and $\Psib=[\psib_1,\cdots, \psib_d] \in \Rd^{d\times d}$ denote the arbitrary basis  matrices
that are multiplied to the left and right
of the Hankel matrix, respectively.
Yin et al.
derived  the following signal expansion,
which they called the {\em convolution framelet} expansion%
~\cite{yin2017tale}:
\begin{eqnarray}\label{eq:frame0}
\x  
&=&  \frac{1}{d} \sum_{i=1}^{n}\sum_{j=1}^d\langle \x,  \phib_i \circledast \psib_j \rangle   \phib_i \circledast \psib_j,
\end{eqnarray}
implying that
$\{\phib_i \circledast \psib_j\}_{i,j=1}^{n,d}$
is a tight frame.}

\add{The observations imply that an efficient and concise signal representation is important for the success of
modern image reconstruction approaches, and specific algorithms may differ in their choice of the frame basis
and specific method to identify the sparse subset that concisely represents the signal.}

\add{In this perspective, the classical approaches such as
CS, structured low-rank Hankel matrix approaches, etc. have two fundamental limitations.
First, the choice of the underlying frame (and its dual)
is based on top-down design principles.
For example, most of wavelet theory has been developed
around the edge-adaptive basis representations
such as  curvelet \cite{candes2006fast}, contourlet \cite{do2005contourlet}, etc.,
whose design principle is based on top-down mathematical modeling.
Moreover, the search for the sparse index set $I$
for the case of compressed sensing
is usually done using a computationally expensive optimization framework.
The following section shows that these limitations
of classical representation learning approaches
can be largely overcome by deep learning approaches.
}

\begin{figure*}[!t]
	\center
	{\includegraphics[width =0.8\textwidth]{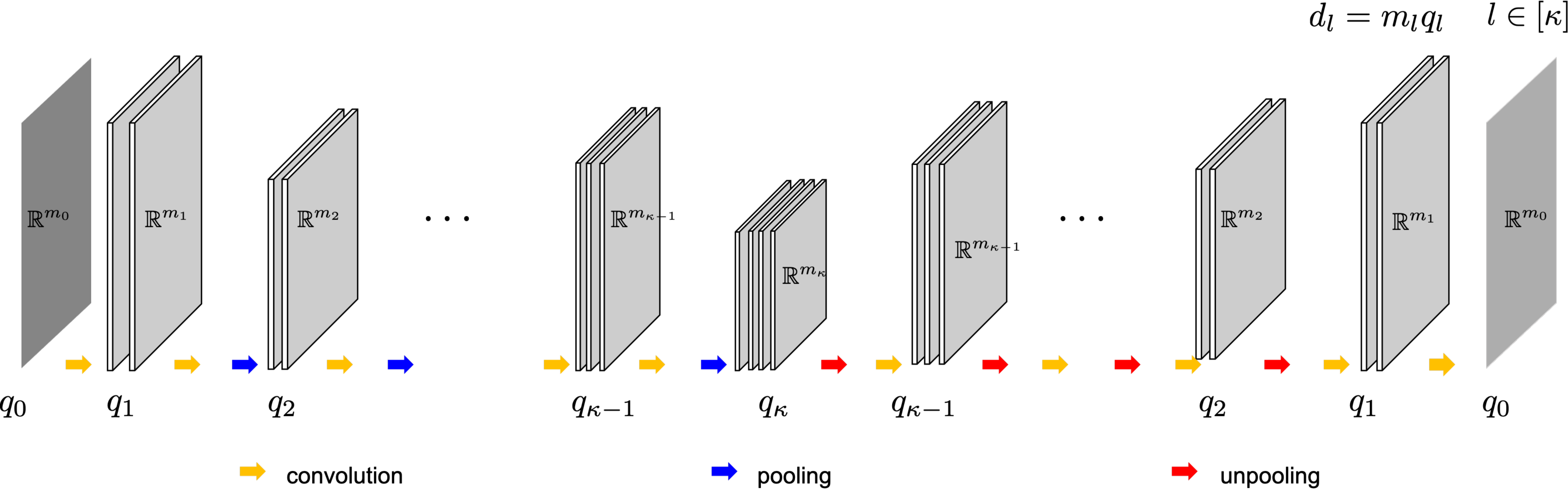}}~~
	\caption{An example of encoder-decoder CNN. \add{The encoder is composed of the first $\kappa$ layers, while  the latter $\kappa$ layers form a decoder network.}  }
	\label{fig:EDCNN}
\end{figure*}

\subsubsection{\add{Deep Neural Networks as Combinatorial Representation Learning}}

The recent theory of deep convolutional framelets
claims that a deep neural network
can be interpreted
as a framelet representation,
whose frame basis is learned from the training data~\cite{ye:18:dcf}.
\add{Moreover, a recent follow-up study~\cite{ye2019understanding} showed how
this frame representation can be automatically adapted to various input signals in a real-time manner.}

To understand these findings, 
consider \add{the} symmetric encoder-decoder CNN  in Fig.~\ref{fig:EDCNN},
which \add{has been} used 
for image reconstruction problems~\cite{jin2017deep,han2017framing}.
Specifically, the encoder network maps a given input signal
$\x\in\Xbc\subset \Rd^{d_0}$
to a  feature space $\z \in \Zbc\subset \Rd^{d_\kappa}$,
whereas the decoder takes this feature map as an input,
processes it,
and produces an output 
$\y \in \Ybc\subset \Rd^{d_0}$.
At the $l$th layer, $m_l$, $q_l$, and $d_l:=m_lq_l$ denote the dimension of the signal,
the number of filter channels,
and the total feature vector dimension, respectively.
We consider a symmetric configuration,
where both the encoder and decoder have the same number of layers, say $\kappa$;
and the encoder layer $\Ec^l$ and the decoder layer $\Dc^l$ are symmetric.
\begin{eqnarray*}
\Ec^l:\Rd^{d_{l-1}} & \mapsto& \Rd^{d_l}, \\ 
\Dc^l:\Rd^{d_{l}} &\mapsto& \Rd^{d_{l-1}}
\end{eqnarray*}
The
$j$th channel output
from the  the $l$th layer encoder
can be represented by a multi-channel convolution operation~\cite{ye2019understanding}:
 \begin{eqnarray}\label{eq:encConv}
\x_j^l = \sigma\left(\Phib^{l\top} \sum_{k=1}^{q_{l-1}}\left(\x_k^{l-1}\circledast \overline \psib_{j,k}^l\right)\right) ,
\end{eqnarray}
where
$\x_k^{l-1}$ denotes the $k$th input channel signal, 
$\overline\psib_{j,k}^l\in \Rd^r$ denotes the $r$-tap convolutional kernel
that is convolved with the $k$th input channel
to contribute to the $j$th channel output,
and $\Phib^{l\top}$ is the pooling operator.
Here, $\overline {\blmath v}$ is  the flipped version of the vector $\blmath v $ such that
$\overline v[n]= v[-n]$ with the periodic boundary condition, and $\circledast$ is the circular convolution. 
(Using periodic boundary conditions
simplifies the mathematical treatments.)
Similarly,
the $j$th channel decoder layer convolution output is given by \cite{ye2019understanding}
  \begin{eqnarray}\label{eq:decConv}
\tilde\x_j^{l-1} = \sigma\left(\sum_{k=1}^{q_{l}}\left(\tilde\Phib^l\tilde\x^{l}_k\circledast  {\tilde\psib_{j,k}^l}\right)\right) , 
\end{eqnarray}
where  $\tilde\Phib^l$ denotes the unpooling operator,
 \add{$\tilde\x_k^{l}$ denotes the $k$th input channel signal for the decoder, and 
$\tilde\psib_{j,k}^l\in \Rd^r$ denotes the $r$-tap convolutional kernel
that is convolved with the $k$th input channel
to contribute to the $j$th channel output.}

By concatenating the multi-channel signal in column direction as
$$
\x^l:=\begin{bmatrix} \x^{l\top}_1 & \cdots & \x^{l\top}_{q_{l}} \end{bmatrix}^\top,
$$
the encoder and decoder convolution in \eqref{eq:encConv} and \eqref{eq:decConv}
can be represented using  matrix notation:
\begin{eqnarray}\label{eq:ED}
 \x^l=\sigma(\Eb^{l\top} \x^{l-1}),& \tilde \x^{l-1}=\sigma(\Db^l \tilde\x^{l})
\end{eqnarray}
where $\sigma(\cdot)$ denotes the element-wise rectified linear unit (ReLU) and 
   \begin{eqnarray}\label{eq:El}
\E^l= \begin{bmatrix} 
\Phib^l\circledast \psib^l_{1,1} & \cdots &  \Phib^l\circledast \psib^l_{q_l,1}  \\
  \vdots & \ddots & \vdots \\
\Phib^l\circledast \psib^l_{1,q_{l-1}} & \cdots &  \Phib^l\circledast \psib^l_{q_{l},q_{l-1}}
 \end{bmatrix}
 \end{eqnarray}
 \begin{eqnarray}\label{eq:Dl}
 \Db^l= \begin{bmatrix} 
\tilde\Phib^l\circledast \tilde\psib^l_{1,1} & \cdots &  \tilde\Phib^l\circledast \tilde\psib^l_{1,q_l}  \\
  \vdots & \ddots & \vdots \\
\tilde\Phib^l\circledast \tilde\psib^l_{q_{l-1},1} & \cdots &  \tilde\Phib^l\circledast \tilde\psib^l_{q_{l-1},q_{l}}
 \end{bmatrix}
 \end{eqnarray}
and
$$\Phib^l=\begin{bmatrix}  \phib^l_1  & \cdots &  \phib^l_{m_l} \end{bmatrix},$$
\begin{eqnarray*}
\begin{bmatrix} \Phib^l \circledast \psib_{i,j}^l  \end{bmatrix}  :=\begin{bmatrix} \phib^l_1 \circledast \psib_{i,j}^l & \cdots & \phib^l_{m_l} \circledast  \psib_{i,j}^l\end{bmatrix}  \label{eq:defconv}
.\end{eqnarray*}
\add{Then, one of the most important observations is that 
the output of the encoder-decoder CNN} can be represented
as follows \cite{ye2019understanding}:
\begin{eqnarray}\label{eq:basis}
\y 
&=& \sum_{i} \langle {\blmath b}_i(\x), \x \rangle \tilde  {\blmath b}_i(\x),
\end{eqnarray}
where $ {\blmath b}_i(\x)$ and $\tilde  {\blmath b}_i(\x)$
denote the $i$th columns of the following frame basis and its dual:
\begin{eqnarray}
\B(\x)&=& \Eb^1\Sigmab^1(\x)\Eb^2 \cdots  \Sigmab^{\kappa-1}(\x)\Eb^{\kappa},~\quad \label{eq:Bc}\\
\tilde \B(\x) &=& \Db^1\tilde\Sigmab^1(\x)\Db^2 \cdots  \tilde\Sigmab^{\kappa-1}(\x)\Db^{\kappa}, \label{eq:tBc}
\end{eqnarray}
and
$\Sigmab^l(\x)$ and $\tilde\Sigmab^l(\x)$ denote diagonal matrices with 0 and 1 values that are determined by the ReLU output
in the previous convolution steps.
Similar basis representation holds for the encoder-decoder CNNs with skipped connection. For more details, see~\cite{ye2019understanding}.

\add{In the absence of ReLU nonlinearities, the authors in \cite{ye:18:dcf,ye2019understanding}
showed that,
assuming that the pooling and unpooling operators
and the filter matrices satisfy appropriate frame conditions for each $l$,
the representation \eqref{eq:basis} is indeed a frame representation of \x 
as in \eqref{eq:Xc},
ensuring perfect signal reconstruction.
However,
in neural networks the input and output should differ,
so the perfect reconstruction condition is not of practical interest.
Furthermore, the signal representation in~\eqref{eq:basis}
should generalize well
for various inputs
rather than for specific inputs at the training phase.}

\add{
Indeed, 
\cite{ye2019understanding} shows that
the explicit  dependence on the input \x in \eqref{eq:Bc} and \eqref{eq:tBc}
due to the ReLU nonlinearity solves the riddle, and
CNN generalizability
comes from the combinatorial nature of the expansion in \eqref{eq:basis}
due to the ReLU.
}

\begin{figure}[ht!]
	\center
	{\includegraphics[width =0.4\textwidth]{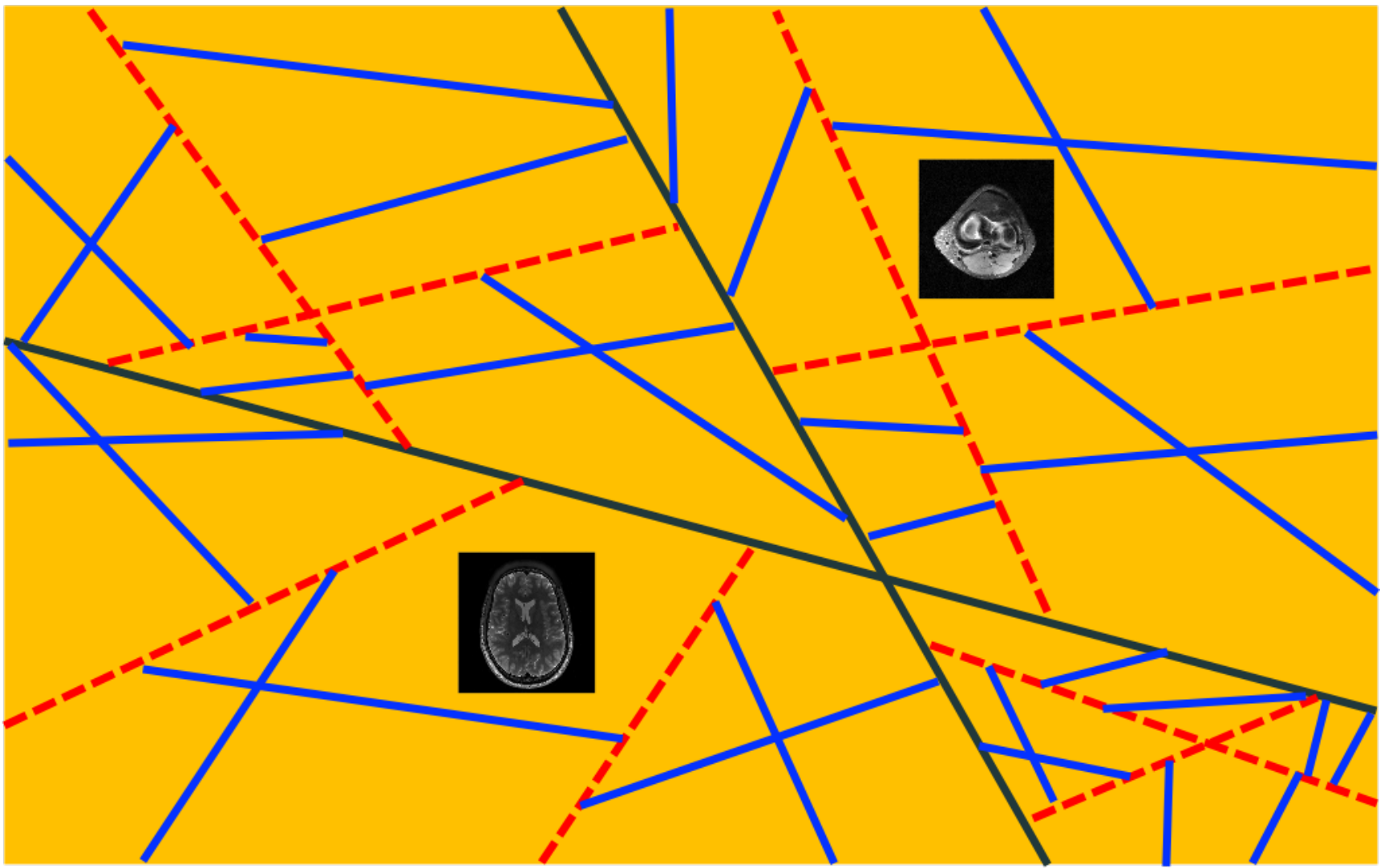}}~~
	\caption{\add{A high level illustration of input space partitioning  for the three layer neural network with two filter channels for each layer. Input images at each partition share the same linear representation, but not across different partitions.}}
	\label{fig:partition}
\end{figure}

\add{Specifically, since the nonlinearity is applied after the convolution operation, the on-and-off  activation pattern of each ReLU
determines a binary partition of the  feature space at each layer across the hyperplane that is determined by the convolution. Accordingly,
 in deep neural networks, 
 the input space $\Xbc$ is partitioned into multiple non-overlapping
regions  as shown in Fig.~\ref{fig:partition} so that input images for each region share the same linear representation, but not accross
the partition.
This implies that
two different input images in Fig.~\ref{fig:partition} are automatically switched   to 
two distinct  linear representations that are different from each other.}

\add{ This  input adaptivity
poses an important computational advantage over the classical representation learning approaches that  rely on computationally expensive optimization techniques.
Moreover, the representations are entirely 
dependent on the filter sets 
that are learned from the training data set, which is different
from the classical  representation learning approaches that are designed by
mathematical principles.
Furthermore, 
the number of input space partitions and the associated distinct linear representations
 increases exponentially
with the network depth, width, and the skipped connection  thanks to the combinatorial
nature of ReLU nonlinearities~\cite{ye2019understanding}. 
This exponentially large  {\em expressivity} of the neural network 
is another important advantage, which may, with the combination of the aforementioned {\em adaptivity},  may explain the origin
of the success of deep neural networks for image reconstruction.}

\section{Open Questions and Future Directions}
\label{sec:open}

There are various challenges, open questions, and directions for image reconstruction
that require further research.
This section
discusses some \add{possible}
directions.

\add{
As background for this section,
it is useful to recall the general topics
of past (and ongoing) image reconstruction research,
many of which are encapsulated
in \eqref{e,xh,map}.
The likelihood $p(\y;\x)$
depends on both
accurate physical modeling
of the imaging system,
and accurate statistical modeling of the measurements.
Numerous papers have focused on improving image quality
by more accurate models for imaging physics or statistics.
Often more accurate models require more computation
so there is an engineering challenge
to make accurate models computationally tractable
for routine use.
Some imaging systems, like MRI,
have quite flexible sampling patterns
and the chosen sampling pattern
is also embedded in the likelihood.
The design of effective sampling patterns
is an active research area in MRI.
The prior $p(\x)$,
or the regularizer $R(\x)$
depends on models for the signal \x.
Numerous papers have focused on this aspect,
and many of the open problems listed below
also relate to signal models.
The ``$\arg \min$''
in \eqref{e,xh,map}
requires an optimization algorithm,
and numerous papers have proposed algorithms (often with various properties such as rapid theoretical or empirical convergence, or whose computations scale well to large-scale settings, etc.)
for specific image reconstruction problems.
This will continue to be an active research area
because new signal models lead to new optimization problems.
Furthermore, many deep-learning methods for image reconstruction
are based on ``unrolling'' some optimization algorithm.}

\add{After performing optimization
to get an image \xh,
the research continues
because the field needs
task-based metrics
to quantify the quality of \xh.
Objective assessment of image quality
is an ongoing research area
\cite{barrett:98:oao}
that should have more influence
on image reconstruction research in the future
because traditional measures like mean-squared error
\cite{wang:09:mse}
are unlikely to predict how well a given imaging system
and reconstruction method
will perform in clinical tasks.
%
}

\add{Several} learning-driven iterative algorithms have been proposed for image reconstruction particularly from limited or corrupted data,
and have shown promise in imaging applications.
Some of these methods have proven convergence guarantees.
For example, recent works~\cite{saibressiam15,sravTCI1}
show the convergence of block coordinate descent transform learning-based blind compressed sensing algorithms to the critical points of the underlying 
nonconvex problems.
However, analysis of theoretical conditions on the learned models,
cost functions, algorithm initializations,
and (e.g., k-space) sampling guaranteeing accurate and stable image recovery
in learning-based setups requires further research.
Such results would shed light on appropriate model properties and constraints
for different modalities and also aid the development of better behaved iterative algorithms.
Theoretical results on desirable properties and invariances
for filters and non-linearities
and provable ways to incorporate physics in the algorithm architecture
would also benefit CNN based reconstruction methods~\cite{leeye18}
and physics-driven deep training-based reconstruction approaches \cite{ravchfess17,chfess18}.

In online learning based reconstruction,
adapting relatively simple models may speed up the algorithm
(particularly when real-time reconstruction is needed),
but at the cost of image quality and vice-versa.
Developing online learning based approaches that achieve optimal 
trade-offs
between complexity or richness of the learned model, runtime per minibatch,
\add{convergence (over time), and image quality} is an important area of future research.

A rigorous understanding of the pros and cons of different learning-based approaches
and the regimes (signal-to-noise ratios, dose levels, or undersampling)
where they work well is lacking.
For example, some methods learn models such as dictionaries or sparsifying transforms
using model-based cost functions from training data.
These methods require fairly modest training data
(e.g., several images or patches),
and can 
learn \add{some} general properties of images
that generalize fairly \add{well~\cite{zheng:18:pua}} to new data
(e.g., unseen anomalies may contain similar directional features). 
Blind compressed sensing methods on the other hand
learn models on-the-fly from measurements without requiring training data,
mimicking multi-layer (iterative) networks
but learned in a completely unsupervised and highly adaptive manner. 
Supervised learning approaches learn the parameters of reconstruction models
often from large datasets of input-output pairs,
but may be less likely to generalize to unseen \add{data~\cite{yeravyongfes18}}
or could produce spurious reconstructions of \add{unseen features~\cite{wensailukebres19}} and anomalies
(which are much less likely to occur in training sets).
Moreover, supervised learning-based methods
typically do not incorporate instance-adaptive components
such as optimizing clustering for each test case within a network.
A rigorous analysis of the different learning methodologies and their efficacy and drawbacks
in different (training and testing) data and noise regimes
would enable better use of such methods
as well as aid the development of better models and improved learning-based reconstruction.
Effectively and efficiently combining the benefits
of both supervised and unsupervised or model-based learning methods
is an interesting line of future research.

There is \add{also} increasing interest in learning-driven sampling of data,
particularly limited measurements, in medical imaging.
Some recent works~\cite{saibresadaptsampl13,learnCSMRI18}
proposed learning the undersampling pattern for CS MRI
to minimize error in reconstructing a set of training images (e.g., pre-scans).
The underlying optimization problems for learning the sampling were combinatorial
and moreover, the reconstruction error that is optimized
would depend on the chosen reconstruction algorithm
and could be a highly nonconvex function as well.
These works~\cite{saibresadaptsampl13,learnCSMRI18}
proposed adapting the sampling to both the training data and the reconstruction algorithm
including learning-based reconstruction schemes,
and showed improved image quality
compared to conventional sampling strategies such as variable density random sampling for MRI.
However, the learning could be computationally very expensive
(e.g., in~\cite{learnCSMRI18})
and the convergence behavior of these sampling adaptation algorithms is unknown.
Development of efficient \add{sampling} 
\add{learning} algorithms with guarantees
would be a promising direction of future research.

\add{
Although we have discussed recent interpretation of deep neural networks
with the perspective of
combinatorial representation learning,
there still remains open questions.
For example, our understanding of the theory
explains input-dependent  automatic adaptation to distinct representation,
but it does not prove whether the adaptation
provides the optimal representation for given input.
This question is closely related
to the generalization power of deep neural networks,
which is still an ongoing research area with many open questions.
Another important open question is that
even if such optimal adaptation from a filter set exists,
our understanding is lacking on
whether the neural network training
can find such optimal filter sets.
Understanding the optimization landscape
is another important research area in deep neural networks,
with many open questions \cite{nguyen2018optimization,ge2017optimization,du2018gradient}.
Recent discoveries show that
in many deep neural networks,
there are many global minimizers during training,
but depending on the optimization algorithms,
different minimizers are selected \cite{gunasekar2018implicit,soudry2018implicit}.
Understanding and exploiting this ``implicit bias''
of the optimization algorithm
is another exciting direction for future research.
}


Finally, given the recent trends and breakthroughs in learning for biomedical imaging,
we expect that the next generation imaging systems
would leverage learning in all aspects of the imaging system.
Such \emph{smart imaging systems} may learn from big datasets
(available locally in hospitals or in the cloud)
as well as from real-time patient inputs
and optimize the sampling for rapid (e.g., with limited measurements) or low-dose imaging,
and also optimize the underlying models
for efficient and effective reconstruction and analytics
(e.g., classification, segmentation, disease feature detection, etc.).
Such adaptation of the data 
acquisition, reconstruction, and analytics components
could be done jointly in an end-to-end manner to maximize performance
in specific clinical tasks
and allowing for both radiologist and patient inputs in the learning process.
The development of these next generation learning-driven systems
would involve research thrusts in both modeling and algorithmic directions
coupled with innovations in physics, hardware, pulse sequence design, etc.
Importantly, we expect models, algorithms, and computation
to play an important and key role in the development of medical imaging in the near future.

\section{Conclusions}
\label{sec:conclusion}
This paper surveyed various advances in the field of medical image reconstruction beginning \add{with analytical} approaches and simple model-based iterative reconstruction methods based on better models of the imaging system physics and sensor statistics and simple image regularization. Then the paper \add{focused on} 
\add{techniques} exploiting improved image models and properties such as sparsity and low-rankness that enable reconstructions from limited or corrupted data, and then \add{discussed several} recent works on sophisticated data-driven or adaptive models and machine learning techniques for reconstruction.
\add{Examples based on specific modalities,}
and discussions were used to provide insight into the behavior and limitations of various \add{types} of surveyed methods.
We discussed the different regimes of adaptivity and learning and some of the connections between different learning-based models and methods.  
While the field of \add{modeling and} \add{learning-based} imaging and the concurrent interest in smart imaging \add{systems is} growing, we discussed some of \add{the} 
challenges, open questions, and future directions for the field in this paper.

\bibliographystyle{ieeetr}
\bibliography{main}




\end{document}